\documentclass[12pt]{article}
\usepackage{amsmath}
\usepackage{amssymb}
\usepackage{bm}
\usepackage{bbm}
\usepackage{cite}
\usepackage{color}
\usepackage{graphicx}
\usepackage{slashed}
\usepackage{tabularx}

\usepackage{epstopdf}
\usepackage[colorlinks]{hyperref}
\hypersetup{
citecolor=blue,
linkcolor=blue,
urlcolor=blue
}

\usepackage{ifthen}
\newcommand{\draftmode}{off}
\ifthenelse{\equal{\draftmode}{on}}{

\setlength{\parindent}{0pt}
\setlength{\parskip}{5ex}

\pagestyle{headings}
\pagestyle{myheadings}
\markright{\hfill \today \hfill}

\usepackage[nofiglist,notablist,nomarkers]{endfloat}
}

\newcommand{\fig}[1]{Fig.~\ref{#1}}

\newcommand{\eq}[1]{Eq.~(\ref{#1})}

\usepackage{subfig}
\newcommand{\GeV}{\mbox{ GeV}}
\newcommand{\MeV}{\mbox{ MeV}}

\begin{document}
\title{\Large\bf
Light mediators in dark matter direct detections
}

\vspace{0.3truecm}
\author{
  Tai Li\footnote{Email: litai@itp.ac.cn},
  Sen Miao\footnote{Email: miaosen@itp.ac.cn}
  and  Yu-Feng Zhou\footnote{Email: yfzhou@itp.ac.cn}
  \\ \\
 \textit{State Key Laboratory of Theoretical Physics},\\
  \textit{Kavli Institute for Theoretical Physics China,}\\
  \textit{Institute of  Theoretical Physics, Chinese Academy of Sciences}\\
  \textit{Beijing, 100190, P.R. China}
}
\date{}
\maketitle

\begin{abstract}
In an extended effective operator framework,
we investigate in detail
the effects of light mediators on
the event spectra of dark matter (DM)-nucleus scatterings.
The presence of light mediators changes
the interpretation of the current experimental data,
especially the determination of DM particle mass.
We show by analytic and numerical illustrations that
in general
for all the  operators relevant to spin-independent scatterings,
the  DM particle mass allowed by
a given set of experimental data
increases significantly when the mediator particle becomes lighter.
For instance,
in the case of CDMS-II-Si experiment,
the allowed DM particle mass  can reach
$\sim 50 \ (100) $~GeV  at  $68\% \ (90\%)$ confidence level,
which is much larger than $\sim 10$ GeV in the case with contact interactions.
The increase of DM particle mass saturates  when
the mediator mass is below $\mathcal{O}(10)$ MeV.
The upper limits from other experiments such as
SuperCDMS, CDMSlite, CDEX, XENON10/100, LUX, PandaX etc.
all tend  to be weaker toward high DM mass regions.
In a combined analysis,
we show that the presence of  light mediators can
partially relax the tension in the current results of  CDMS-II-Si,
SuperCDMS and LUX.
\end{abstract}

\newpage

\section{Introduction}
Weakly interacting massive particles (WIMPs) are
popular candidates for dark matter (DM) which
contributes to $26.8\%$ of the energy ÊÊbudget  of the Universe~\cite{Ade:2013zuv}.
At present
numerous underground DM direct detection experiments are
searching for possible signals arising from
the interactions between WIMPs and the  Standard Model~(SM) particles.

In the recent years,
a number of  experiments such as
DAMA~\cite{Bernabei:2008yi,Bernabei:2010mq},
CoGeNT~\cite{Aalseth:2011wp,Aalseth:2012if,Aalseth:2014eft},
and
CDMS-II-Si~\cite{Agnese:2013rvf}
have reported possible signals in excess of known backgrounds.
Other experiments, such as
CDMS-II-Ge~\cite{Akerib:2010pv,Ahmed:2010wy},
CDMSlite~\cite{Agnese:2013jaa},
SuperCDMS~\cite{Agnese:2014aze},
XENON10~\cite{Angle:2011th},
XENON100~\cite{Aprile:2012nq},
LUX~\cite{Akerib:2013tjd},
PandaX~\cite{Xiao:2014xyn},
SIMPLE~\cite{Felizardo:2014awa}, and
CDEX~\cite{Yue:2014qdu}, etc.
only reported  upper limits on the scattering cross section.
The recent results of CDMS-II-Si,
if interpreted in terms of
DM-nucleus elastic scattering in simple benchmark  DM models,
favor a DM particle mass  $\sim 8.5$ GeV
and a  spin-independent DM-nucleon scattering cross section
$\sim 2\times 10^{-41}~\rm{cm}^{2}$,
which is in tension with the limits of other experiment such as
SuperCDMS and LUX.
Note that
in the interpretation of the experimental data,
the interactions between DM particles and target nuclei are
often  assumed to be contact, elastic, isospin conserving,
momentum and velocity independent, etc..
Simplified assumptions are also adopted on
the DM velocity distribution, DM local energy density,
nuclear form factors, detector responses, etc..
The interpretation of the  experimental data can be
changed dramatically  if
any of the assumptions is modified.

In some DM models
the interaction between
DM particle and target nuclei is of long-range type through
the exchange of light mediator particles.
The mediator can be  the SM photon
if the DM particle carries
tiny effective electric charge
\cite{
Goldberg:1986nk,
Cheung:2007ut,
Feldman:2007wj
},
or electric/magnetic dipole moments
\cite{Pospelov:2000bq,
Sigurdson:2004zp,
Gardner:2008yn,
Heo:2009vt,
Masso:2009mu,
Barger:2010gv,
Chang:2010en,
Cho:2010br,
Banks:2010eh,
Fortin:2011hv,
Kumar:2011iy,
Barger:2012pf,
DelNobile:2012tx,
Weiner:2012cb},
or anapole moment
\cite{Ho:2012bg,
Fitzpatrick:2010br,
Frandsen:2013cna,
Gresham:2013mua}.
Another example of the mediator particle is
a massive dark photon,
a hidden sector $U(1)$ gauge boson
interacting with the SM particles through
kinetically mixing with the SM photon
( for  reviews, see e.g.
Refs.~\cite{Hewett:2012ns,
Jaeckel:2013ija,
Essig:2013lka} ).
The mediator can also be  a scalar particle
interacting with SM fermions
\cite{Schwetz:2011xm,
Fornengo:2011sz,
Cotta:2013jna,
Curtin:2013qsa,
An:2013xka}.
The recently observed excess in cosmic-ray positron fraction  by
PAMELA
\cite{Adriani:2008zr,
Adriani:2010ib},
Fermi-LAT
\cite{FermiLAT:2011ab}
and
AMS-02
\cite{Accardo:2014lma},
if interpreted  in terms of halo DM annihilation,
requires a DM annihilation cross section significantly larger
than the typical thermal DM annihilation cross section
(see  e.g. \cite{Jin:2013nta,Jin:2014ica}).
The presence of light mediators can  lead to
long-range attractive forces between DM particles,
which enhance the DM annihilation cross sections
at low temperatures through
the mechanism of Sommerfeld enhancement
\cite{
Sommerfeld,Hisano:2002fk,Hisano:2003ec,Cirelli:2007xd,ArkaniHamed:2008qn,
Pospelov:2008jd,MarchRussell:2008tu,Iengo:2009ni,Cassel:2009wt,
Chen:2013bi,
Liu:2013vha
}.
Furthermore,
the DM self-interactions with light mediator mass  around
$\mathcal{O}(1-100)$~MeV
can help in understanding the problems in
small-scale structure (core vs. cusp) of  the DM profile of
dwarf galaxies
\cite{Loeb:2010gj,
Tulin:2013teo,
Kaplinghat:2013yxa
}.

In the light mediator scenario,
the spectra of the recoil event rate
are expected to be enhanced at
low momentum transfer,
which depends on both the type of the interaction and
the mass of the mediator.
The DM interpretation of the experimental data can
be significantly different from the case of  contact interactions.
In this work,
we investigate the effects of light mediator on
the interpretation of the current experimental data,
especially on
the determination of DM particle mass
in an extended effective operator framework
\cite{Chang:2009yt,Fan:2010gt,Fitzpatrick:2012ix,Kumar:2013iva}.
According to the momentum and velocity dependencies,
the effective operators are catalogued into six types.
Using analytical and numerical illustrations,
we show that in general
the allowed DM particle mass from the experimental data
increases significantly when the mediator particle becomes lighter.
For the data of CDMS-II-Si,
the allowed DM particle mass  can reach
$\sim 50 \ (100)$~GeV at 68\% (90\%) confidence level (CL),
depending on the type of operators,
and
the increase  saturates  when
the mediator mass is below $\mathcal{O}(10)$ MeV.
The upper limits from other experiments such as
SuperCDMS, CDMSlite, CDEX, XENON10/100, LUX, PandaX etc.
all become weaker toward high DM mass region.
At present,
the most stringent limits are from the results of SuperCDMS and LUX,
which can be marginally consistent with that  of CDMS-II-Si in
DM models with xenonphobic  interactions.
We perform a combined analysis on
the data of CDMS-II-Si, SuperCDMS and LUX
and
show that the presence of a light mediator can
partially relax the tension in the current results of  CDMS-II-Si,
SuperCDMS and LUX.

This  paper is organized as follows.
In Section~\ref{sec:The-rate},
we overview
the extended  effective operator approach in
the presence of  light mediators,
and
discuss the momentum and velocity dependences of the
differential cross sections for all the operators.
In Section~\ref{sec:mediator-mass},
we discuss the effect of light mediator on the determination of DM properties,
especially the mass of DM particles.
In Section~\ref{sec:Direct-detection-data},
we discuss the interpretation of the current experimental data of
CDMS-II-Si, SuperCDMS, LUX and other experiments.
The conclusions are given in Section~\ref{sec:Conclusion}.

\section{Effective operators with a light mediator particle\label{sec:The-rate}}
In many DM  models,
the  DM particle $\chi$ can scatter  off a target nucleon $N=(p,n)$ in
an elastic process
$\chi(\boldsymbol{p}_{1})+N(\boldsymbol{p}_{2})\to \chi(\boldsymbol{p}_{3})+N(\boldsymbol{p}_{4})$ via exchanging a mediator particle $\phi$ in $t$-channel.
For fermionic DM particles,
if the mass of the mediator $m_{\phi}$ is much larger than
the  3-momentum transfer $q=|\mathbf{q}|=|\boldsymbol{p}_{3}-\boldsymbol{p}_{1}|$ of
the scattering process, i.e., $m_{\phi}^{2} \gg q^{2}$,
the interactions  responsible for the scattering  can be
effectively described by a set of local  Lorentz-invariant operators
\begin{align}
\mathcal{O}_{i}=\frac{c_{i}}{\Lambda^{2}}
(\bar\chi \Gamma_{i} \chi) (\bar N \Gamma_{i}' N) ,
\end{align}
where $\Lambda$ is the mass scale of the mediator particle,
and $c_{i}$ are the coefficients.
The matrices $\Gamma_{i},\Gamma'_{i}$ are Lorentz-invariant combinations of
the Dirac matrices
$$
\Gamma_{i},\Gamma'_{i}=\{ 1,\ \gamma^{5},\ \gamma^{\mu},\ \gamma^{\mu}\gamma^{5},\
\sigma^{\mu\nu},\ \sigma^{\mu\nu}\gamma^{5}\}.
$$

When the mediator is relatively light,
the correction to the effective operator approach can be made  by
a replacement $\Lambda^{2} \to (q^{2}+m_{\phi}^{2})$,
arising from the $t$-channel $\phi$-exchange.
For fermionic DM particles,
up to dimension six,
the operators with  scalar  and  pseudo-scalar interactions are
\begin{align}\label{eq:fermion}
\mathcal{O}_{1}& = \frac{1}{q^{2}+m_{\phi}^{2}}\bar{\chi}\chi\bar{N}N, \nonumber
&\mathcal{O}_{2}& = \frac{1}{q^{2}+m_{\phi}^{2}}\bar{\chi}\gamma^{5}\chi\bar{N}N,
\\
\mathcal{O}_{3} &= \frac{1}{q^{2}+m_{\phi}^{2}}\chi\chi\bar{N}\gamma^{5}N,
&\mathcal{O}_{4} &=\frac{1}{q^{2}+m_{\phi}^{2}}\bar{\chi}\gamma^{5}\chi\bar{N}\gamma^{5}N .
\end{align}
The operators with vector or axial-vector interactions are
\begin{align}
\mathcal{O}_{5}  &=  \frac{1}{q^{2}+m_{\phi}^{2}}\bar{\chi}\gamma^{\mu}\chi\bar{N}\gamma_{\mu}N,\nonumber
&\mathcal{O}_{6}  &=  \frac{1}{q^{2}+m_{\phi}^{2}}\bar{\chi}\gamma^{\mu}\gamma^{5}\chi\bar{N}\gamma_{\mu}N,
\\
\mathcal{O}_{7}  &=  \frac{1}{q^{2}+m_{\phi}^{2}}\bar{\chi}\gamma^{\mu}\chi\bar{N}\gamma_{\mu}\gamma^{5}N,
&\mathcal{O}_{8}&  =  \frac{1}{q^{2}+m_{\phi}^{2}}\bar{\chi}\gamma^{\mu}\gamma^{5}\chi\bar{N}\gamma_{\mu}\gamma^{5}N,
\end{align}
and that with tensor interactions are
\begin{align}
\mathcal{O}_{9}=\frac{1}{q^{2}+m_{\phi}^{2}}\bar{\chi}\sigma^{\mu\nu}\chi\bar{N}\sigma_{\mu\nu}N,  \quad
\mathcal{O}_{10}=\frac{1}{q^{2}+m_{\phi}^{2}}\bar{\chi}\sigma^{\mu\nu}\gamma^{5}\chi\bar{N}\sigma_{\mu\nu}N .
\end{align}
Note that other combinations of the tensor operators are not independent  due to the identity
$\sigma^{\mu\nu}\gamma^{5}=(i/2)\epsilon^{\mu\nu\alpha\beta}\sigma_{\alpha\beta}$.
If the DM particles are Majorana,
the vector and tensor operators are vanishing identically.
Similarly,
if the DM particles are complex scalars,
possible operators  up to dimension six are
\begin{align}\label{eq:complexScalar}
\mathcal{O}_{11}  &=  \frac{2m_{\chi}}{q^{2}+m_{\phi}^{2}}\chi^{*}\chi\bar{N}N,\nonumber
&\mathcal{O}_{12}&  =  \frac{2m_{\chi}}{q^{2}+m_{\phi}^{2}}\chi^{*}\chi\bar{N}\gamma^{5}N,
\\
\mathcal{O}_{13} &=  \frac{1}{q^{2}+m_{\phi}^{2}}(\chi^{*}\overleftrightarrow{\partial_{\mu}}\chi)\bar{N}\gamma^{\mu}N,
&\mathcal{O}_{14}& =  \frac{1}{q^{2}+m_{\phi}^{2}}(\chi^{*}\overleftrightarrow{\partial_{\mu}}\chi)\bar{N}\gamma^{\mu}\gamma^{5}N.
\end{align}
For real scalar DM particle,
the vector operators $\mathcal{O}_{13,14}$ vanish identically.
Throughout  this work,
for simplicity,
we assume that
one of the operators dominates the scattering processes at a time.
It is straight forward to extend the analysis to
the processes involving  multiple operators simultaneously.

The differential cross section for $\chi N$ scattering can be written as
\begin{equation}
\frac{d\sigma_N}{d q^2}(q^2,v)=\frac{  \overline{\left|\mathcal{M}_{\chi N}\right|^{2} }}{64\pi m_{N}^{2}m_{\chi}^{2}v^{2}},
\end{equation}
where
$\overline{\left|\mathcal{M}_{\chi N}\right|^{2}}$ is
the squared  matrix element averaged over the spins of  initial states,
and $v=|\boldsymbol{v}|=|\boldsymbol{p}_{1}/m_{\chi}-\boldsymbol{p}_{2}/m_{N}|$ is the
velocity of the DM particle in the nucleon rest frame.
In general,
the differential cross section $d\sigma_{N}/dq^2$ depends on both  $q^{2}$ and $v$,
and can be divergent in the limit of $q^{2}\to 0$,
when the mass of the mediator is vanishing.
The total DM-necleon scattering cross section $\sigma_{N}$ are defined as
the integration of $d\sigma_{N}/dq^2$ over $q^{2}$
in the interval from an infrared  cutoff $q^{2}_{\text{min}}$ to
the maximal value allowed by kinematics $q_{\text{max}}^{2}=4\mu_{\chi N}^{2}v^{2}$
\begin{equation}
\sigma_{N}(v)
=
\int_{q^{2}_{\text{min}}}^{4\mu_{\chi N}^{2}v^{2}} d q^2 \frac{d\sigma_{N}}{dq^2}(q^{2},v),
\label{eq:sigmaNdef}
\end{equation}
where $\mu_{\chi N}=m_{\chi}m_{N}/(m_{\chi}+m_{N})$ is the DM-nucleon reduced mass.
The value of $q^{2}_{\text{min}}$ can be related to
the energy threshold of DM detection experiment which
is typically at keV scale.

Since $\sigma_{N}(v)$ is in general a function of $v$,
it is useful to  define a velocity-independent cross section $\bar\sigma_{N}$ which is
the total cross section at a reference velocity $v_{\text{ref}}$, i.e.,
$\bar\sigma_{N}\equiv\sigma_{N}(v_{\text{ref}})$.
The value of $v_{\text{ref}}$ can be chosen to be
the typical velocity of halo DM particles
$\sim 200~\text{km}\cdot\text{s}^{-1}$.
Thus the differential cross section  can be rewritten in the conventional form
\begin{equation}
\frac{d\sigma_N}{d q^2}(q^2,v) \equiv \frac{\bar\sigma_N}{4\mu_{\chi N}^2 v^2} G(q^2,v),
\end{equation}
where
\begin{equation}
G(q^2,v)
=\frac{(q_{\text{ref}}^{2}-q_{\text{min}}^{2})
\overline{\left|\mathcal{M}_{\chi N}(q^2,v)\right|^{2}} }
{\intop_{q^{2}_{\text{min}}}^{q_{\text{ref}}^{2} } dq^2
\overline{\left|\mathcal{M}_{\chi N}(q^2,v_{\text{ref}})\right|^{2}} },
\label{eq:G_N}
\end{equation}
is a factor containing  the $q^{2}$-dependence and the rest of $v$-dependence,
and $q_{\text{ref}}^{2} \equiv 4\mu_{\chi N}^{2}v^{2}_{\text{ref}}$.
It is clear that $G(q^{2},v)= 1$, when the matrix element $M_{\chi N}$ is a constant.

The DM models often contribute directly to
one of a combination of
the above mentioned Lorentz-invariant relativistic operators.
In the nonrelativistic limit,
the matrix element $M_{\chi N}(q^{2},v)$ can be expressed
in terms of the products of
independent Galilean invariant vectors such as
the 3-momentum transfer $\boldsymbol{q}$,
the transverse velocity of the DM particle
$\boldsymbol{v}_{\bot}\equiv\boldsymbol{v}+ \boldsymbol{q}/(2\mu_{\chi N})$,
and
the spin of DM particle (nucleon) $\boldsymbol{S}_{\chi}$ ($\boldsymbol{S}_{N}$).
An important property of the transverse velocity is that it is perpendicular to $\boldsymbol{q}$
in elastic $\chi N$ scatterings, $\boldsymbol{v}_{\bot}\cdot \boldsymbol{q}=0$
and in this case the value of $v_{\bot}$ is given by
\begin{align}\label{eq:vodot}
v^{2}_{\bot}=v^{2}- \frac{q^{2}}{4\mu^{2}_{\chi N}} .
\end{align}
Detailed discussions related to the properties of  $\boldsymbol{v}_{\bot}$ and nonrelativistic
operators can be found in Refs.\cite{Fitzpatrick:2012ix,Kumar:2013iva}.
In this work,
we focus on the spin-independent cross sections which
involve  $q^{2}$ and $v^{2}_{\bot}$ only.
According to the $v_{\bot}^{2}$ and $q^{2}$ dependences,
the operators in Eqs.~(\ref{eq:fermion})--(\ref{eq:complexScalar}) can be
catalogued into  the following six types:

\noindent {\bf Type-I operators}
$\mathcal{O}_{1}$,  $\mathcal{O}_{5}$, $\mathcal{O}_{11}$ and $\mathcal{O}_{13}$ correspond to scalar or vector type interactions.
These operators contribution to the matrix elements with the form
\begin{align}
|M_{\chi N}(q^{2},v)|^{2}
\propto
\frac{1}{(q^{2}+m_{\phi}^{2})^{2}}.
\end{align}
For type-I operators, the factor $G(q^2,v)$ is
\begin{align}
G_{1}(q^{2})
=
\frac{1}
{I_{1}\left(
q^{2}_{\text{min}}/m^{2}_{\phi}, q_{\text{ref}}^{2}/m^{2}_{\phi}
\right)
\left(1+q^{2}/m_{\phi}^{2}\right)^{2}
}  ,
\end{align}
where
the function $I_{1}(a,b)$ is defined as
\begin{align}
I_{1}(a,b)
\equiv
\frac{1}{b-a}\int^{b}_{a} dt \frac{1}{(1+t)^{2}}
=\frac{1}{(1+a)(1+b)} .
\end{align}
In the limit
\begin{align}\label{eq:heavylimit}
q^{2}_{\text{min}} \ll q_{\text{ref}}^{2}\ll m_{\phi}^{2},
\end{align}
the function $I_1$ can be approximated by
$I_{1}\approx 1-q_{\text{ref}}^{2}/m_{\phi}^{2}$.
It is evident that for a heavy mediator, $q_{\text{ref}}^{2}, q^{2} \ll m^{2}_{\phi}$,
$G_{1}(q^{2})\approx 1$, as expected.

\noindent{\bf Type-II operators}
$\mathcal{O}_{2}$ and  $\mathcal{O}_{10}$
contribute to  the matrix elements of the form
\begin{align}
|M_{\chi N}(q^{2},v)|^{2}
\propto
\frac{q^{2}}{(q^{2}+m_{\phi}^{2})^{2}} .
\end{align}
For type-II operators,
the corresponding factor is
\begin{align}
G_{2}(q^{2})
=
\frac{q^{2}/m_{\phi}^{2}}
{I_{2}
\left(
q^{2}_{\text{min}}/m^{2}_{\phi}, q_{\text{ref}}^{2}/m^{2}_{\phi}
\right)
\left(1+q^{2}/m_{\phi}^{2}\right)^{2}
} ,
\end{align}
where
\begin{align}
I_{2}(a,b)
\equiv
\frac{1}{b-a}\int^{b}_{a} dt \frac{t}{(1+t)^{2}}
=
\frac{1}{b-a}\ln\left( \frac{1+b}{1+a}  \right)-
I_{1}(a,b) .
\end{align}
In the limit  of \eq{eq:heavylimit},
the function $I_{2}$ can be approximated by
$I_{2}\approx q_{\text{ref}}^{2}/(2 m_{\phi}^{2})$.
Thus in the case of a heavy mediator,
$G_2(q^2)$ can be reduced to
$G_2(q^2) \approx  2 q^2/q_{\text{ref}}^2$ which
is suppressed by the smallness of $q^{2}$.

\noindent
{\bf Type-III operator} $\mathcal{O}_{6}$ corresponds to the anapole type of
interactions which results in the matrix elements of the form
\begin{align}
|M_{\chi N}(q^{2},v)|^{2}
\propto
\frac{v_{\bot}^{2}}{(q^{2}+m_{\phi}^{2})^{2}}
\end{align}
For type-III operators, the factor $G_{3}(q^2,v)$ is given by
\begin{align}
G_{3}(q^{2},v)
=
\frac{
v_{\bot}^{2}/v^{2}_{\text{ref}}}
{I_{3}
\left(
q^{2}_{\text{min}}/m^{2}_{\phi}, q_{\text{ref}}^{2}/m^{2}_{\phi}
\right)
\left(1+q^{2}/m_{\phi}^{2}\right)^{2}
}
\end{align}
where
$I_{3}(a,b)=I_{1}(a,b)-I_{2}(a,b)/b$.
In the limit  of \eq{eq:heavylimit},
the function $I_{3}$ can be approximated by
$I_{3}\approx 1/2-q_{\text{ref}}^{2}/(3 m_{\phi}^{2})$.
Thus in the heavy mediator case,
$ G_{3}(q^2,v) \approx 2 v^{2}_{\bot}/v^{2}_{\text{ref}}$.

The operators
$\mathcal{O}_{7}$, $\mathcal{O}_{8}$ and $\mathcal{O}_{9}$
contribute dominantly to spin-dependent cross sections.
Their subdominant contributions to
the spin-independent cross sections can be
catalogued into three types (type-IV, -V and -VI).
They are suppressed by higher powers of $q^{2}$ or $v_{\bot}^{2}$.
Since these operators are
now severely  constrained by the null-results from
PICASSO~\cite{Archambault:2012pm} and
COUPP~\cite{Behnke:2012ys}, etc.,
they are not considered
in the remainder of this work. 
The factors  $G(q^{2},v)$ for these operators are shown in  the Appendix.
Note that the operators
$\mathcal{O}_{3}$, $\mathcal{O}_{4}$
$\mathcal{O}_{12}$ and $\mathcal{O}_{14}$
give no contribution to
the spin-independent cross section.

For the operators under consideration,
at nucleus level,
the spin-independent DM-nucleus differential cross section
$d\sigma_A/dq^2$ can be written as follows
\begin{equation}
\frac{d\sigma_{A}}{dq^{2}}
=
 \frac{\bar \sigma_{p}}{4\mu_{\chi p}^{2}v^{2}}\left[Z+\xi (A-Z)\right]^{2}G(q^2,v)F_{A}^{2}(q^2),
\label{eq:dsigmaAdq2}
\end{equation}
where $Z$ and $A$ are the atomic number and atomic mass number of the target nucleus,
respectively,
and $\xi$ is the relative strength between DM-neutron and DM-proton coupling.
The case where $\xi \neq 1$ corresponding to the isospin violating DM,
and for $\xi=-0.7$, the XENON100 constraints can be weakened maximumly, see e.g.
\cite{
Feng:2011vu,
Frandsen:2011ts
}.
Note that large isospin violating is subject to constraints from  the data of
cosmic-ray photons
\cite{arXiv:1112.4849}
and antiprotons
\cite{Jin:2012jn,Zhou:2012qv}.
The factor $F_{A}(q^{2})$ is related to the distribution of
the nucleon within the nucleus.
For type-I, type-II and type-IV operators,
it is simply the form factor
$F_{A}(q^{2})$=$F(q^{2})=\int d^{3}x \rho(x) e^{ iqx}$ where
$\rho(x)$ is  the density distribution function of the nucleon inside the nucleus.
However,
for type-III, type-V and type-VI operators
because the nucleon velocity operator $\boldsymbol{v}$ acting on
the nucleus wave function will pick up the nucleus mass,
the reduced mass $\mu_{\chi N}$ in the expression of
$v_{\bot}$ in \eq{eq:vodot} will be replaced by $\mu_{\chi A}$
(for detailed derivation, see e.g.~\cite{Fitzpatrick:2012ix}).
This effect can be effectively absorbed into the expression of $F_{A}(q^{2})$ as follows
\begin{equation}
F_{A}(q^{2})=\frac{v^{2}-q^{2}/4\mu_{\chi A}^{2}}{v^{2}-q^{2}/4\mu_{\chi p}^{2}} F(q^{2}) .
\end{equation}
The form factor $F^2(q^2)$ can be taken as  the Helm form\cite{Lewin:1995rx},
\begin{equation}
F^2(q^2)= \frac{9 j_{1}^{2}(q R_{1})}{q^2 R^2_{1} } e^{-q^2 s^2},
\end{equation}
where $j_{1}(x)$ is the spherical Bessel function, $R_{1}=\sqrt{R_{A}^{2}-5s^{2}}$
with $R_{A}\simeq1.2A^{1/3}$ fm is an effective nuclear radius and $s\simeq1$ fm.

\section{Determination of  DM particle mass}\label{sec:mediator-mass}

For  GeV scale WIMP DM particles,
the typical momentum transfer $q$ in the DM-nucleus scattering process is
at MeV scale.
In the case where
the value of  $m^{2}_{\phi}$ is comparable with $q^{2}$,
the predicted spectrum of $dR/dE_{R}$ is significantly different from
the case with contact interactions
and
the mass of the DM particle determined from
fitting to the measured  $dR/dE_{R}$ becomes
sensitive to the value of $m_{\phi}$.
The differential recoil event rate per detector mass is given by
\begin{equation}\label{eq:eventRate}
\frac{dR}{dE_{R}}
=
\frac{dN}{M_{A} dt dE_{R}}
=
\frac{2 \rho_{\chi}}{m_{\chi}}
\int_{|\boldsymbol{v}|>v_{\text{min}}}d^{3}\boldsymbol{v}vf(\boldsymbol{v})\frac{d\sigma_{A}}{dq^{2}},
\end{equation}
where
$E_{R}=q^{2}/(2m_{A})$ is the nuclear recoil energy,
$\rho_{\chi} \approx 0.3\text{ GeV}\cdot\text{cm}^{-3}$ is the local DM energy density,
and
$M_{A}$ is the total mass of the target nucleus (in units of kg).
The minimal velocity  required to
generate the recoil energy $E_R$ in elastic scatterings is
$v_{\text{min}}=\sqrt{m_{A} E_{R}/(2\mu_{\chi A}^{2} )}$
with $\mu_{\chi A}=m_{\chi}m_A/(m_{\chi}+m_A)$  the DM-nucleus reduced mass.
The DM velocity distribution function $f(\boldsymbol{v})$ is related to
that in the Galactic frame $f_{G}(\boldsymbol{v})$ through
a Galilean transformation
$f(\boldsymbol{v})=f_{G}(\boldsymbol{v}+\boldsymbol{v}_{E};v_{0},v_{\text{esc}})$,
where
$\boldsymbol{v}_{E}$ is the velocity of the Earth relative to the rest frame of the Galactic halo,
$v_{0}\approx 220 \text{ km}\cdot\text{s}^{-1}$ is
the most probable velocity of the DM particle,
and
$v_{\text{esc}}\approx 544\text{ km}\cdot\text{s}^{-1}$ is
the Galactic escape velocity from the solar system.
The DM velocity distribution in the Galactic halo rest frame is
often assumed to be
the standard Maxwell-Boltzmann distribution with
a cutoff at $v_{\text{esc}}$,
\begin{equation}
f_{G}(\boldsymbol{v})
=
\frac{1}{N}
\exp \left( -\frac{v^{2}}{v_{0}^{2}}\right)
\Theta(v_{\text{esc}}-v) ,
\label{eq:fv}
\end{equation}
with the normalization constant  given by
$N=(\pi v_{0}^{2})^{3/2}\text{erf}(v_{\text{esc}}/v_{0})-2v_{\text{esc}}\text{exp}(-v_{\text{esc}}^{2}/v_{0}^{2})/(\pi v_0^2)$.
It is useful to define the  slope of the recoil event spectrum as follows
\begin{equation}\label{eq:appr_der_rate}
S \left( q^2\right) \equiv - 2 m_A \frac{d^2R/dE_R^2}{dR/dE_R},
\end{equation}
which in principle can be determined by experiment,
if the statistics  of the recoil events is high enough.

For the operators under consideration,
the expressions of the recoil event rates involve
two type of velocity  integrals
\begin{align}\label{eq:g1}
g_{1}(v_{\text{min}})
&=\int_{v_{\text{min}}}\frac{f(\boldsymbol{v})}{v}d^{3}{\boldsymbol v}
\nonumber \\
&=\frac{1}{2v_{E}}\left[{\rm erf}\left(\frac{v_{+}}{v_{0}}\right)-{\rm erf}\left(\frac{v_{-}}{v_{0}}\right)\right]
-\frac{1}{\sqrt{\pi}v_{E}}\left(\frac{v_{+}-v_{-}}{v_{0}}\right) e^{-v^2_{\text{esc}}/v_{0}^{2}}
\end{align}
and
\begin{align}\label{eq:g2}
g_{2}(v_{\text{min}})
=& \int_{v_{\text{min}}}vf(\boldsymbol{v})d^{3}{\boldsymbol v}
\nonumber \\
=& \frac{v_{0}}{\sqrt{\pi}}\left[ \left(\frac{v_{-}}{2 v_E} + 1 \right)
e^{-v^2_{-}/v_{0}^{2}}
- \left(\frac{v_{+}}{2 v_E} - 1 \right) e^{-v^2_{+}/v_{0}^{2}}\right]
\nonumber\\
& + \frac{v_{0}^{2}}{4v_{E}}\left(1+2\frac{v_{E}^{2}}{v_{0}^{2}}\right)\left[{\rm erf}\left(\frac{v_{+}}{v_{0}}\right)-{\rm erf}\left(\frac{v_{-}}{v_{0}}\right)\right]
\\
&-  \frac{v_{0}}{\sqrt{\pi}}\left[2+\frac{1}{3v_{E}v_{0}^{2}}\left(\left(v_{\text{min}}+v_{\text{esc}}-v_{-}\right)^{3}-\left(v_{\text{min}}+v_{\text{esc}}-v_{+}\right)^{3}\right)\right]
e^{-v^2_{\text{esc}}/v_{0}^{2}},\nonumber
\end{align}
where $v_{\pm}=\text{min}(v_{\text{min}}\pm v_{E},v_{\text{esc}})$.

In general,
the spectral feature of the differential event rate depends on
both $m_{\chi}$ and $m_{\phi}$.
Thus  the value of $m_{\chi}$ can be determined as a function of $m_{\phi}$.
In order to illustrate of the role of $m_{\phi}$ in the determination of DM particle mass,
we first consider an extreme case where
$v_E \ll v_{\text{min}}$ and $v_{\text{esc}}\rightarrow \infty$.
In this limit,
$f(\boldsymbol{v})\approx f_{G}(\boldsymbol{v})$
and
$g_{1,2}(v_{\text{min}})$ can be approximated by
\begin{align}\label{eq:appr_g}
g_{1}(v_{\text{min}})&\approx \frac{2}{\sqrt{\pi}v_0 }e^{-v^2_{\text{min}}/v_{0}^{2}}
\ \text{ and } \
g_{2}(v_{\text{min}})  \approx  \frac{2}{\sqrt{\pi}v_0 }\left( v^2_{\text{min}}+v_0^2 \right) e^{-v^2_{\text{min}}/v_{0}^{2}}.
\end{align}
Neglecting the $q^{2}$-dependence in the form factor $F(q^{2})$,
for type-I operators,
the differential event rate is given by 
\begin{equation}
\left( \frac{dR}{dE_{R}} \right)_{\text{type-I}}
\approx
\frac{2C}{\sqrt{\pi}v_0}
\frac{1}{(q^2+m^2_{\phi })^2}
\exp\left(-\frac{q^{2}}{4\mu^2_{\chi A}v_0^2}\right) .
\label{eq:appr_rate_I}
\end{equation}
where
$C=\rho_\chi  \bar\sigma_p [Z+\xi (A-Z)]^2 /(2 m_\chi \mu^2_{\chi p})$
is a constant independent on $q^{2}$.
For type-I operators, $S(q^{2})$ is  nearly a constant, if $m_{\phi}^{2} \gg q^{2}$.
Let us define a quantity  $S_{\text{exp}}(q^{2})$ which is assumed to be the value of $S(q^{2})$ measured by the experiment at a momentum transfer $q^{2}$.
For a given  $S_{\text{exp}}(q^{2})$,
the value of $m_\chi$ can be determined as a function of $m_{\phi}$  as follows
\begin{equation}
m_{\chi}=\frac{m_A}{ v_0 \sqrt{S_{\text{exp}}(q^{2})-\frac{8 m_A^2}{q^2+m^2_\phi}} - 1}.
\label{eq:relationI}
\end{equation}
The dependences of  $m_{\chi}$ on $m_{\phi}$ have
different  behaviour in three parameter regions:
\begin{itemize}
\item[A)]
in the region where  $8 m^{2}_{A}/(q^{2}+m_{\phi}^{2}) \ll S_{\text{exp}}(q^{2})$,
the determined value of $m_{\chi}$ is almost insensitive to $m_{\phi}$.
\item[B)]
in the region where
the values of $8 m^{2}_{A}/(q^{2}+m_{\phi}^{2})$ and $S_{\text{exp}}(q^{2})$ are
of the same order of magnitude,
the  value of $m_{\chi}$  increases with a decreasing $m_{\phi}$.
When $m_{\phi}^{2}$ is vanishing,
the value of $m_{\chi}$ saturates and reaches its maximal value
\begin{equation}
m_{\chi,\text{max}}
=
\frac{m_A}{v_0 \sqrt{S_{\text{exp}}(q^{2})-8m_A^2/q^2}-1} .
\end{equation}
\item[C)]
in a special parameter region where
$S_{\text{exp}}(q^{2})-8 m^{2}_{A}/(q^{2}+m_{\phi}^{2}) \approx 1/v^{2}_{0}$,
the value of $m_{\chi}$ becomes very large and even divergent.
Thus in this case,
the mass of the DM particle is not constrained  by the data.
\end{itemize}
\begin{figure}\begin{center}
\includegraphics[width=0.3\textwidth]{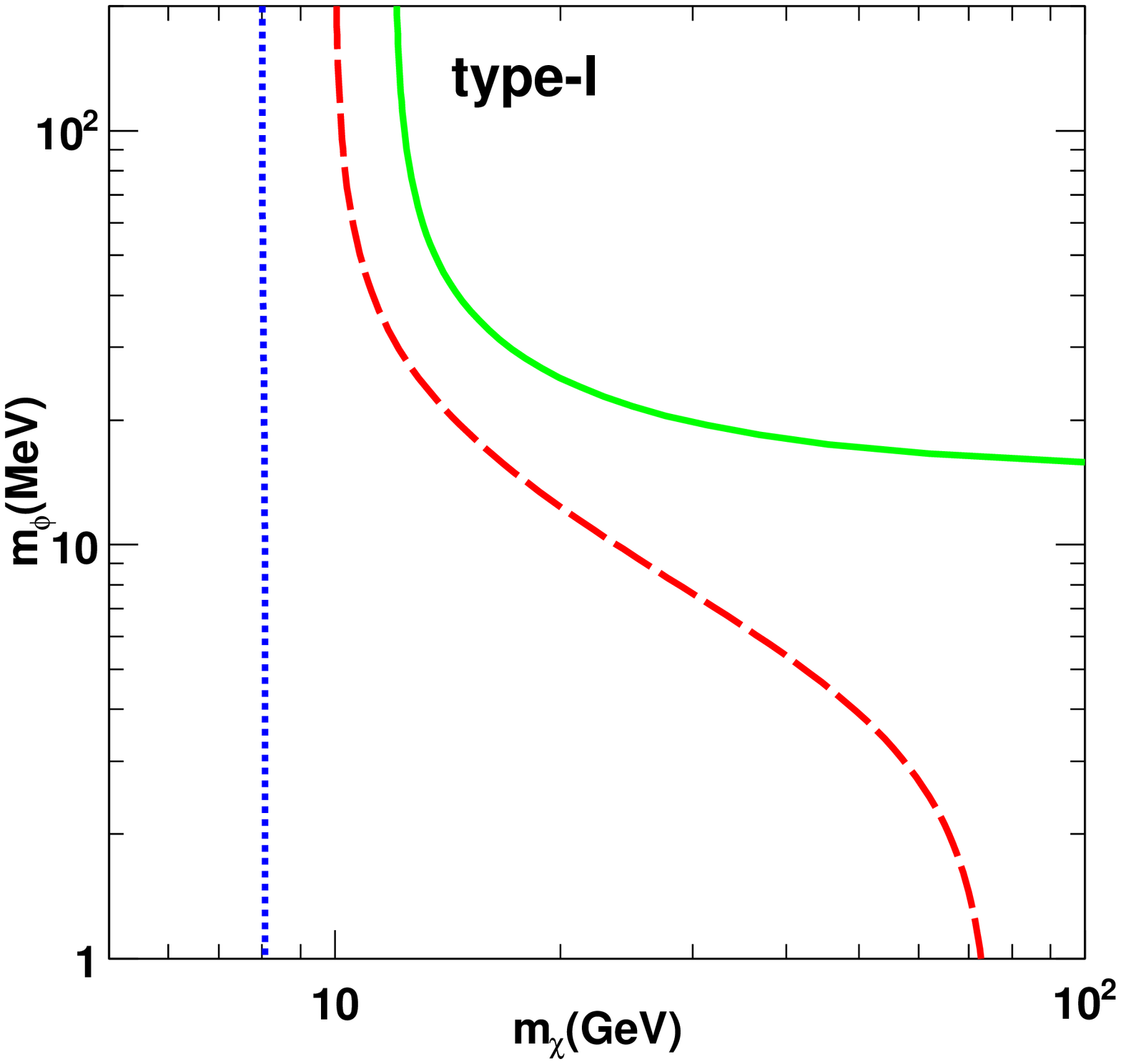}
\includegraphics[width=0.3\textwidth]{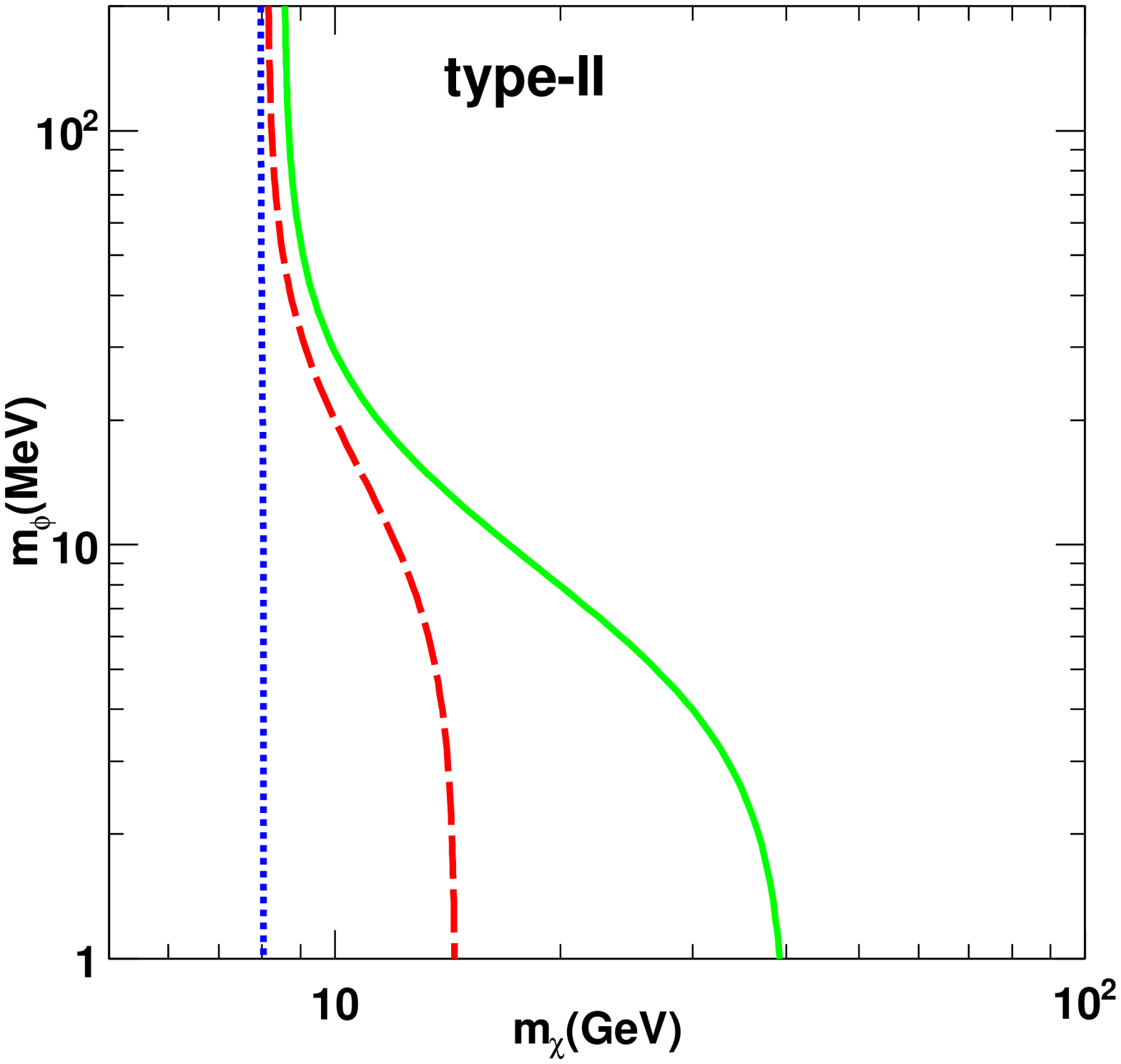}
\includegraphics[width=0.3\textwidth]{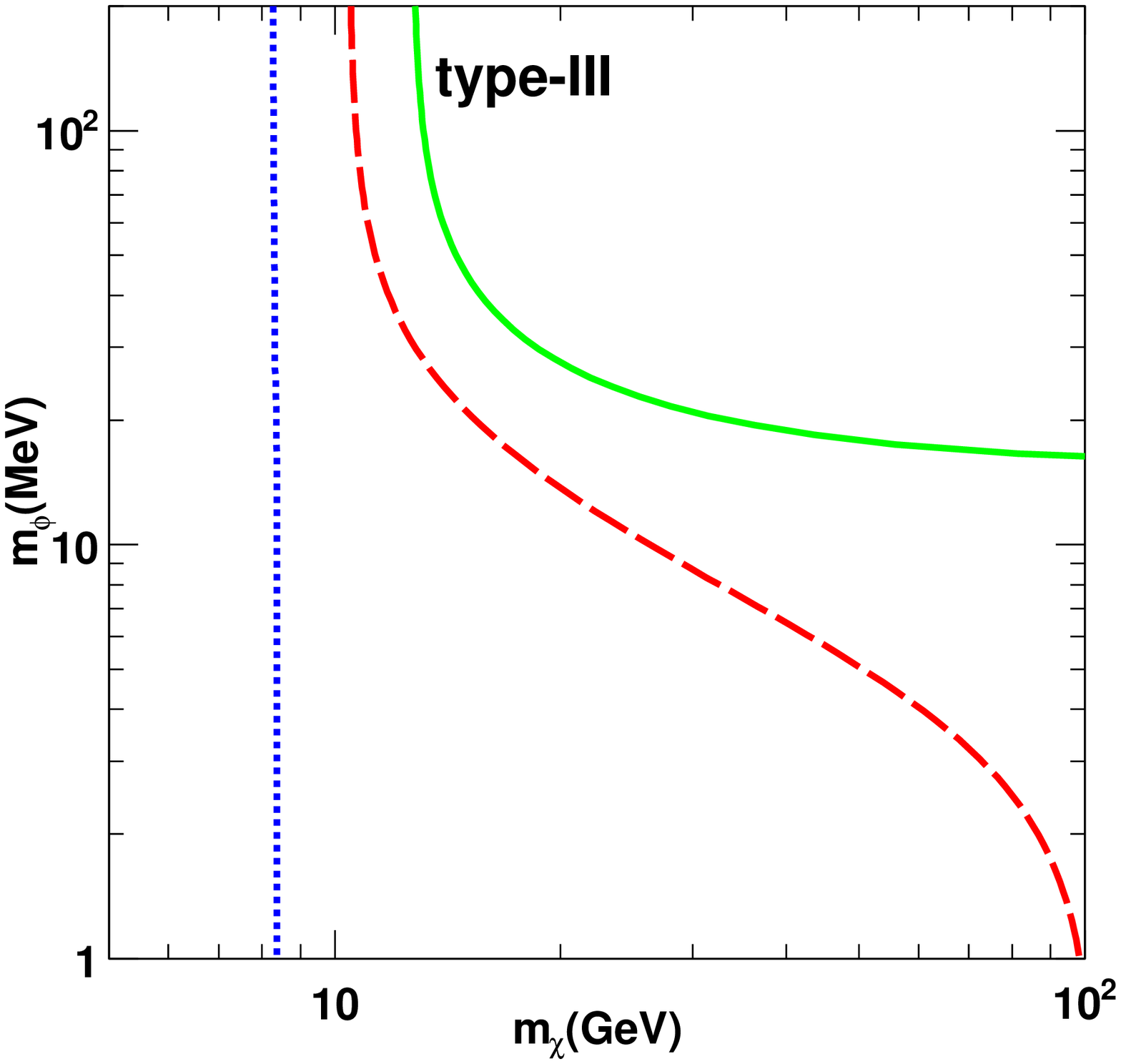}

\caption{
\label{fig:examples}
Relations between $m_\chi$ and $m_\phi$ for
three sets of values of
$S_{\text{exp}}$ and $q^{2}$ listed  in \eq{eq:Sexp}
for type-I, -II and -III operators.
The three curves in each panel corresponds to the case
A (blue dotted), B (red dashed) and C (green solid) in \eq{eq:Sexp},
respectively.
The target nucleus is assumed to be silicon.
}
\label{fig:examples}
\end{center}
\end{figure}

For Type-II operators,
the differential event rate is given by
\begin{equation}
\left( \frac{dR}{dE_{R}} \right)_{\text{type-II}}
\approx \frac{2C}{\sqrt{\pi}v_0}
\left( \frac{q^2}{q^{2}_{\text{ref}}} \right)
\frac{1}{(q^2+m^2_{\phi })^2}
\exp\left(-\frac{q^{2}}{4\mu^2_{\chi A}v_0^2}\right)
.
\label{eq:appr_rate_II}
\end{equation}

Similarly,
given the value of $S_{\text{exp}}(q^{2})$, $m_\chi$ as a function of $m_\phi$ can be obtained as
\begin{equation}
m_{\chi}=\frac{m_A}{ v_0 \sqrt{S_{\text{exp}}(q^{2})-\frac{8m_A^2}{q^2+m^2_\phi}+\frac{4m_A^2}{q^2}} - 1}.
\label{eq:relationII}
\end{equation}
Thus for the same values of $S_{\text{exp}}(q^{2})$ and $m_\phi$, Type-II operators lead to a smaller $m_\chi$ compared with Type I operators.

For Type-III operators,
the integration over velocity results in a simple relation
\begin{equation}\label{eq:typerelation}
\int_{v_{\text{min}}}^{\infty} d^{3}\boldsymbol{v}
\left(\frac{v^{2}_{\bot}}{v^{2}_{\text{ref}}}\right)
\frac{f_{G}(\boldsymbol{v})}{v}
=\frac{2}{\sqrt\pi v_{0}}
\left(\frac{v_{0}^{2}}{v_{\text{ref}}^{2}} \right)
\exp\left(
-\frac{q^{2}}{4 \mu_{\chi A}v_{0}^{2}}
\right),
\end{equation}
which shows that in the limit $f(\boldsymbol{v})\approx f_{G}(\boldsymbol{v})$,
the differential event rate has the same $q^{2}$ dependence as that
in type-I operators.
Consequently,
\begin{equation}\label{eq:typerelation2}
\left( \frac{dR}{dE_{R}} \right)_{\text{type-III}}
\approx
\left(\frac{v_{0}^{2}}{v_{\text{ref}}^{2}}\right)
\left( \frac{dR}{dE_{R}} \right)_{\text{type-I}}
\end{equation}
Thus the relation between $m_\chi$ and $m_\phi$ are the same as that for
type-I operators in Eq.~(\ref{eq:relationI}) in the limit of
$v_{E}\ll v_{\text{min}}$ and $v_{\text{esc}}\to \infty$.

In  more realistic cases where $v_{E}$ is comparable with $v_{\text{min}}$ in size,
the full expressions of $g_{1,2}(v_{\text{min}})$ in
Eqs.(\ref{eq:g1})-(\ref{eq:g2}) have to be applied,
and the value of the DM particle mass can  be determined numerically.
For a concrete  illustration,
we assume that the target nucleus is made of silicon.
We consider three reference values of $S_{\text{exp}}(q^{2})$,
which are chosen  according to  the three typical cases
A, B and C discussed previously.
\begin{align}\label{eq:Sexp}
A) \quad & S_{\text{exp}}=5.25\times 10^{7},   \ \text{at} \ q^{2}=900~\text{MeV}^{2} ,
\nonumber\\
B) \quad & S_{\text{exp}}=1.57\times 10^{7},  \ \text{at} \ q^{2}=400~\text{MeV}^{2} ,
\nonumber\\
C) \quad & S_{\text{exp}}=1.13\times 10^{7},  \ \text{at} \ q^{2}=324~\text{MeV}^{2},
\end{align}
and solve the value of $m_{\chi}$ from \eq{eq:appr_der_rate} for as a function of  $m_{\phi}$.
The velocity of the Earth is set to be
$v_{E}\approx v_{\odot}$
with
$v_{\odot}=v_{0}+12 \text{ km}\cdot\text{s}^{-1}$ being the velocity of the Sun in the Galactic halo rest frame.
Thus the effect of solar modulation is neglected.
In \fig{fig:examples},
the solutions for $m_{\chi}$ for three type of operators are shown.
It can be seen that the three reference values of $S_{\text{exp}}$ roughly
correspond to the three cases of A, B and C discussed in  the limit of
$v_{E}\ll v_{\text{min}}$ and $v_{\text{esc}}\to \infty$.
Comparing the results for three different types of operators,
one sees that the type-II operators indeed lead to a smaller $m_\chi$ than
type-I operators for a given $m_\phi$.
For type-III operators,
although the result is quite similar to the case of type-I operators,
it  predicts  a slightly larger $m_\chi$.
In  DM direct detection experiments,
the target nuclei are of different type
and
the measured spectra can be in a wide range of $q^2$ (or $E_R$).
However,
the observation that a smaller $m_\phi$ can lead to a larger $m_\chi$  is
quite general,
as it will be shown in the next section.

\section{
Interpretation of the  direct detection data  in the light mediator scenario
}\label{sec:Direct-detection-data}

In the DM direct detection experiments,
the measured event signal $s$ could be
the electron-recoil equivalent energy $E_{\text{ee}}$,
direct scintillation signal S1, and
ionization electron charge signal S2, etc..
The differential signal event rate is given by
\begin{equation}
\frac{dR}{ds}
=
\int_{0}^{\infty}
dE_{R} \varepsilon(s) P(s,E_{R})
\frac{dR}{dE_{R}},\label{eq:spectrum}
\end{equation}
where
$P(s,E_{R})$ is the possibility of  observing a signal $s$,
given a recoil energy $E_{R}$.
For a given experiment,
the signal $s$ is related to $E_{R}$ through  a  function $s=f(E_{R})$.
For a detector with perfect signal resolution,
$P(s,E_{R})=\delta(s-f(E_{R}))$.
The efficiency of detecting the signal $s$ is denoted by $\varepsilon(s)$.
In the case where
$P(s,E_{R})$ is Gaussian and $\varepsilon$ is a constant,
the expected total event number in a  signal range $[s_1,$ $s_2]$ is given by
\begin{align}\label{eq:rateResulution}
N_{[s_{1},s_{2}]}=\text{Ex} \int^{\infty}_{0}
dE_{R}
\left(\frac{dR}{dE_{R}}\right)
\frac{\varepsilon}{2}
\left[
	\frac{{\rm erf}(s_{1}-f(E_{R}))}{\sqrt{2}\sigma}
	-\frac{{\rm erf}(s_{2}-f(E_{R}))}{\sqrt{2}\sigma}
\right] ,
\end{align}
where $\text{Ex}=M_{A} \Delta t$ is the exposure,
and $\sigma$ is the signal resolution.

In this section,
we constrain the parameters related to the DM properties
such as
$m_\chi$, $m_\phi$, $\xi$ and $\sigma_p$ from
the experimental data through
evaluating  the function
$\chi^2= - \sum 2 \rm ln \mathcal L $
with $\mathcal{L}$ being the likelihood function.
The likelihood function is chosen according to
the extended maximum likelihood method which
takes into account the distribution of the signal events for
the experiments coming with unbinned data~\cite{Barlow:1990vc}
\begin{equation}
\mathcal{L}\label{eq:EML}
=
e^{-(N+B)}
\prod_{i}^{n}
\left[
\left(\frac{dN}{ds}\right)_{i}+
\left(\frac{dB}{ds}\right)_{i}
\right],
\end{equation}
where
$N$($B$) is the expected total number of signal events from
DM (background) in the whole measured recoil energy range.
The corresponding differential event rate
at the $i$-th event ($i=1,\dots,n$) is denoted by $(dN(B)/ds)_{i}$.
We first vary the values of parameters to
obtain the minimal value of the $\chi^{2}$ function $\chi^{2}_{\text{min}}$,
then calculate $\Delta \chi^{2}=\chi^{2}-\chi^{2}_{\text{min}}$
which is assumed to follow a $\chi^{2}$ distribution.
The allowed regions of parameter space at
68\%, 90\% and 95\%  CL
correspond to  $\Delta \chi^{2}=2.3$, 4.6 and 6.0, respectively,
for the case with two degrees of freedom (d.o.f).

\subsection{The experimental data}
At present,
the most stringent constraints on the spin-independent
cross sections come from the data of
SuperCDMS~\cite{Agnese:2014aze} and
LUX~\cite{Akerib:2013tjd},
which are in strong disagreement  with the positive results
from DAMA~\cite{Bernabei:2008yi,Bernabei:2010mq}
and
CoGeNT~\cite{Aalseth:2011wp,Aalseth:2012if,Aalseth:2014eft}.
The CoGeNT result  is also challenged by  that from the CDEX experiment
which utilizes the same type of germanium  detector~\cite{Yue:2014qdu},
and
there are still debates on the uncertainties in the analysis of the  CoGeNT data
~\cite{Aalseth:2014jpa,Davis:2014bla}.
Only the positive signals from
CDMS-II-Si~\cite{Agnese:2013rvf}
can be marginally consistent with
the limits from LUX and SuperCDMS in
fine tuned DM models with xenonphobic or Ge-phobic  interactions.
Thus in this work,
we  shall mainly focus on the interpretation and compatibility of
the  following three experiments:
\begin{itemize}
\item {\bf CDMS-II-Si},
the CDMS-II experiment measures both
the ionization electrons and non-equilibrium phonons.
Recently, the CDMS-II experiment reported
an observation of 3 possible DM-induced events  with
recoil energies at $E_{R}$=8.2, 9.5  and 12.3 keV,
respectively,
in its silicon detectors,
based on a raw exposure of 140.2 kg$\cdot$days~\cite{Agnese:2013rvf}.
The estimated background from surface event is
$0.41^{+0.20}_{-0.08}(\text{stat.})^{+0.28}_{-0.24}(\text{syst.})$
and
that from neutrons and $^{206}\text{Pb}$ are
$<0.13$ and $<0.08$ at the $90\%$ CL, respectively.
We take  $B=0.62$ as a conservative estimate
and assume a constant $dB/dE_{R}$ which is normalized to
give the total event number $B$ in the recoil energy interval 7--100 keV.
The acceptance efficiency is obtained from Fig.~1 of Ref.~\cite{Agnese:2013rvf}.
For the CDMS-II-Si experiment, a perfect energy resolution is assumed.

\item {\bf SuperCDMS},
with an exposure of 577 kg$\cdot$days,
the  SuperCDMS experiment observed $N=11$ nuclear recoil  events
in the energy range of 1.6--10.0 keV~\cite{Agnese:2014aze}.
The estimated total background is  $B=6.1^{+1.1}_{-0.8}$ events.
The recoil energies of the 11 events were listed in Tab.~1 of Ref.~\cite{Agnese:2014aze},
and
the  acceptance efficiency $\varepsilon(E_{R})$ is obtained from
the Fig.~1 of Ref.~\cite{Agnese:2014aze}.
Again a perfect energy resolution is assumed for the SuperCDMS experiment.

\item {\bf LUX},
the LUX experiment utilizes
a dual-phase XENON time-projection chamber which
measures both the prompt scintillation signal S1 and ionization electron charge signal S2.
With an exposure of $1.01\times 10^{4}$ kg$\cdot$days,
only one candidate event at S1=3.2 PE ($\sim 4.9$ keV) marginally passed the selection cuts
in the S1 signal range $2-30$ PE~\cite{Akerib:2013tjd}.
The estimated number of background is $B=0.64\pm0.16$.
The event number of S1 signals in a given interval $[\text{S1}_{a}, \text{S1}_{b}]$
is given by
\begin{equation}\label{eq:LUX}
N=\int_{\text{S1}_{a}}^{\text{S1}_{b}} d\text{S1}
\sum_{s1=1}^{\infty}\varepsilon(\text{S1})
\text{Gauss}(\text{S1}|s1,\sigma)
\int_{0}^{\infty}
\text{Poiss}(s1|f(E_{R}))\varepsilon_{\text{S2}}(E_{R})\frac{dR}{dE_{R}}dE_{R},
\end{equation}
where $\sigma$ is the energy resolution.
For the LUX experiment
$\sigma=\sqrt{s1}\sigma_{\text{PMT}}$ with $\sigma_{\text{PMT}}=0.37$~PE.
We read off the S1 detection efficiency $\varepsilon(\text{S1})$ from Fig.~1 of Ref.~\cite{Akerib:2013tjd},
and $f(E_{R})$  from the S1-$\log_{10}$(S2/S1)  plots in Fig.~3 and 4 of Ref.~\cite{Akerib:2013tjd}.
The candidate event at S1=3.2 PE corresponds to a recoil energy $E_{R} \simeq 4.9$ keV.
An additional S2 efficiency cutoff $\varepsilon_{\text{S2}}$  for $E_R<3.0$~keV is adopted
~\cite{Gresham:2013mua,DelNobile:2013gba}.

\end{itemize}

For a comparison purpose,
we also calculate upper limits from the following experiments.
Unless otherwise stated, the  limits  are obtained  using the maximum gap method~\cite{Yellin:2002xd}.
\begin{itemize}

\item {\bf CDMSlite}.
The exposure of CDMSlite is 6.3 kg$\cdot$days.
The recoil energy spectrum is shown in Fig. 1 of Ref.~\cite{Agnese:2013jaa}.
We consider the energy range from 0.17 to 7.00 keVee,
and use the histogram with bin-width of  10 eVee for
the energy spectrum from 0.10 to 1.60 keVee,
and the histogram with bin-width of 75 eVee for energy range from 1.60 to 7.00 keVee.
The  detection efficiency is taken to be 98.5\% and the energy resolution is 0.014 keVee.
The recoil energy $E_R$ is reconstructed from
the electron-recoil equivalent energy  $E_{\text{ee}}$ the same as defined in Ref.~\cite{Agnese:2013jaa}.
Since the derivation of upper limits from maximum gap method requires
requires unbinned data,
Following Ref.~\cite{DelNobile:2013gba},
we rearrange the data set by dividing each bin with multiple events into
smaller equal-size bins so that there is one event in each bin.

\item {\bf CDEX}
the CDEX-1 experiment utilizes a  P-type point-contact germannium detector.
The DM search based on a 53.9 kg$\cdot$days of exposure did not
show excess of events over the expected background,
and
stringent upper limits on DM-nucleus scattering cross section were obtained
~\cite{Yue:2014qdu}.
For CDEX, we construct a Gaussian likelihood function
\begin{equation}
\mathcal{L}
=
\prod_{i}
\frac{1}{\sqrt{2\pi}\sigma_{i,\text{exp}}}
\exp\left[
	\frac{(R_{i,\text{exp}}-R_{i,\text{DM}})^{2}}{2\sigma_{i,\text{exp}}^{2}}
	\right],
\end{equation}
where for bin $i$, $R_{i,\text{exp}}$ and $\sigma_{i,\text{exp}}$ are the detected event rate and error per keVee-kg-day,
which is read off from Fig.~3(b) of Ref.~\cite{Yue:2014qdu},
$R_{i,\text{DM}}$ is the theoretical predicted event rate from DM plus a free constant background.
The quenching factor is taken from Ref.~\cite{Lin:2007ka},
and a perfect energy resolution is assumed.

\item {\bf XENON100},
with an exposure of 225 days $\times$ 34 kg,
2 events were reported as the possible excess,
with recoil energies of 7.1 keV~(3.3 PE) and 7.8 keV~(3.8 PE)~\cite{Aprile:2012nq}.
The background expectation is $B=1.0$ events.
The expected spectrum are also calculated using \eq{eq:LUX},
where we take
$f(E_R)=(S_{\text{nr}}/S_{\text{ee}}) L_y E_R \mathcal{L}_{\text{eff}} (E_R)$,
with
$S_{\text{nr}}=0.95$,
$S_{\text{ee}}=0.58$, and
$L_y=2.28$
\cite{Aprile:2012nq}.
We adopt the measurements of
$\varepsilon(\text{S1})$ and $\mathcal{L}_{\text{eff}}$ from
Ref.~\cite{Aprile:2013teh}.
For XENON100 experiment,  $\sigma_{\text{PMT}}=0.5$ PE.
An S2 efficiency cutoff $\varepsilon_{\text{S2}}$ for $E_R<3.0$ keV is adopted.

\item{\bf XENON10},
we use the results of an S2-only analysis on XENON10~\cite{Angle:2011th}
with an exposure of 15 kg$\cdot$days.
The recoil energies of the candidate events are read  off from Fig. 2 of Ref.~\cite{Angle:2011th}.
The expected event number in a given energy range [$E_1,$ $E_2$] is
calculated using \eq{eq:rateResulution},
where the energy resolution is $\sigma=E_R/\sqrt{E_R Q_y}$.
The electron yield $Q_y$ is read off from the solid curve in Fig.~1 of Ref.~\cite{Angle:2011th},
as that used by the XENON10 collaboration.
A cutoff that $Q_y$ vanishes for $E_R<1.4$ keV is adopted.
In the numerical calculation, we adopt a flat efficiency of $\varepsilon=0.94$.

\item {\bf PandaX}
is an other dual-phase xenon time-projection chamber located at CJPL.
The first DM search results from PandaX-1  were based on
a 643.8 kg$\cdot$days of exposure, and
no DM candidate event was found.
In this work, the expected spectrum of PandaX is also calculated using Eq.~(\ref{eq:LUX}).
The efficiency cut $\varepsilon(\text{S1})$ is read off from
the black curve (the overall NR detection efficiency) in Fig. 3 of Ref.~\cite{Xiao:2014xyn}.
The values of
$f(E_R)$ are read off from the red solid curve in Fig. 5 of Ref.~\cite{Xiao:2014xyn}.
Furthermore,
the upper limits are obtained  assuming a Poisson distribution of DM induced events with
an expected background of $B=0.15$ events.

\end{itemize}

%

\subsection{Results}
\begin{figure}
\begin{center}
\subfloat[]{\includegraphics[scale=0.27]{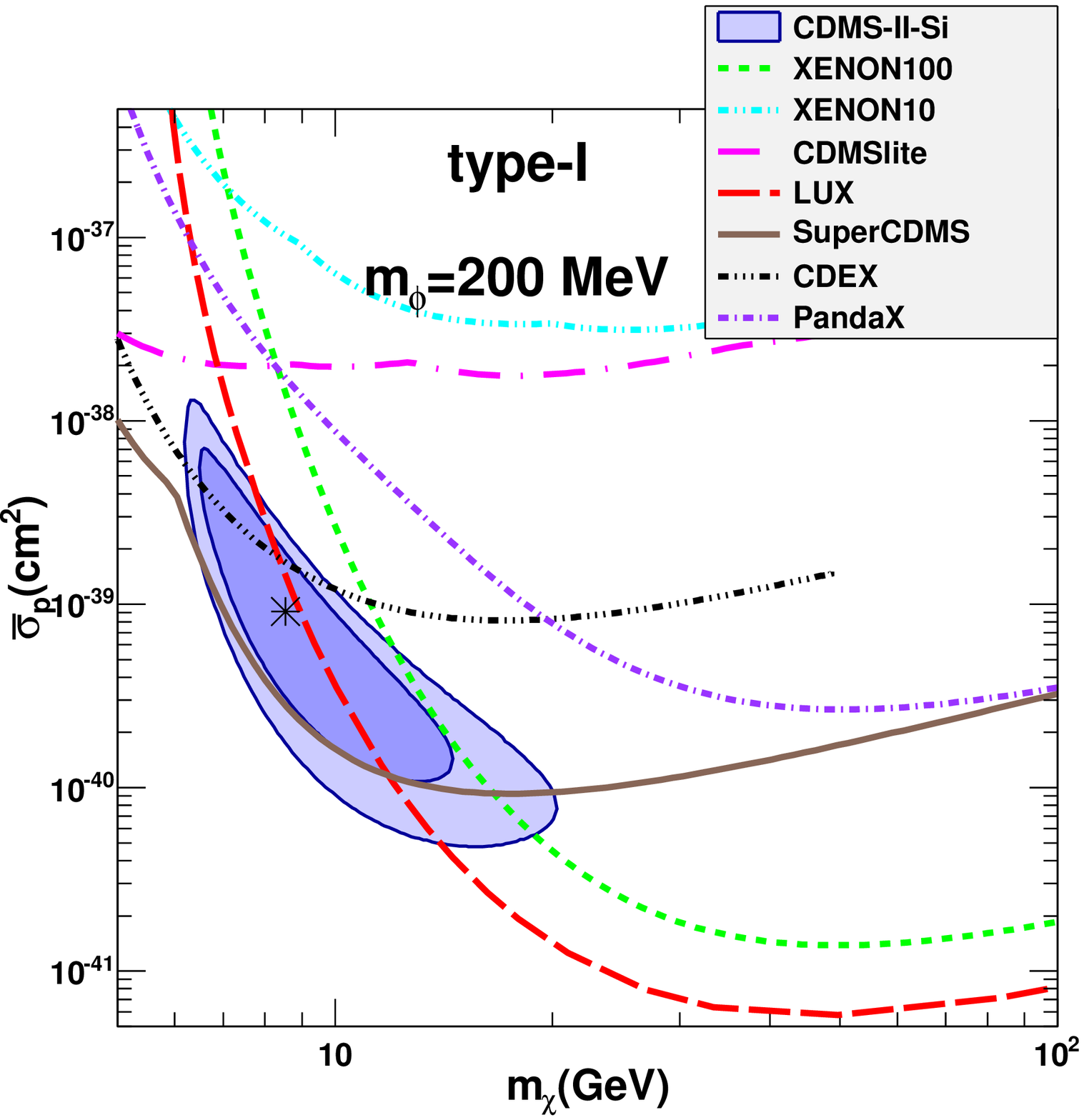}}
\subfloat[]{\includegraphics[scale=0.27]{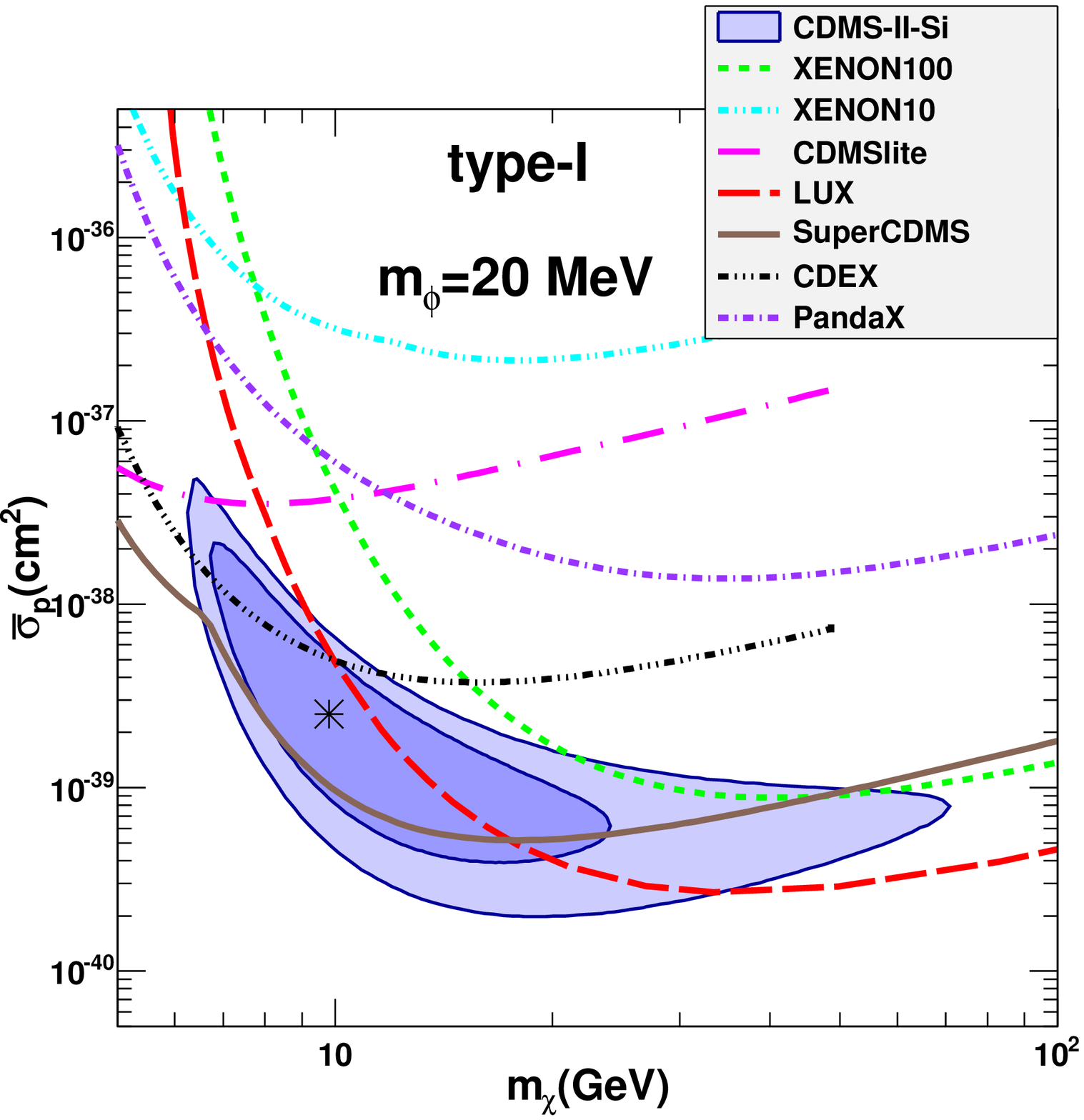}}
\subfloat[]{\includegraphics[scale=0.27]{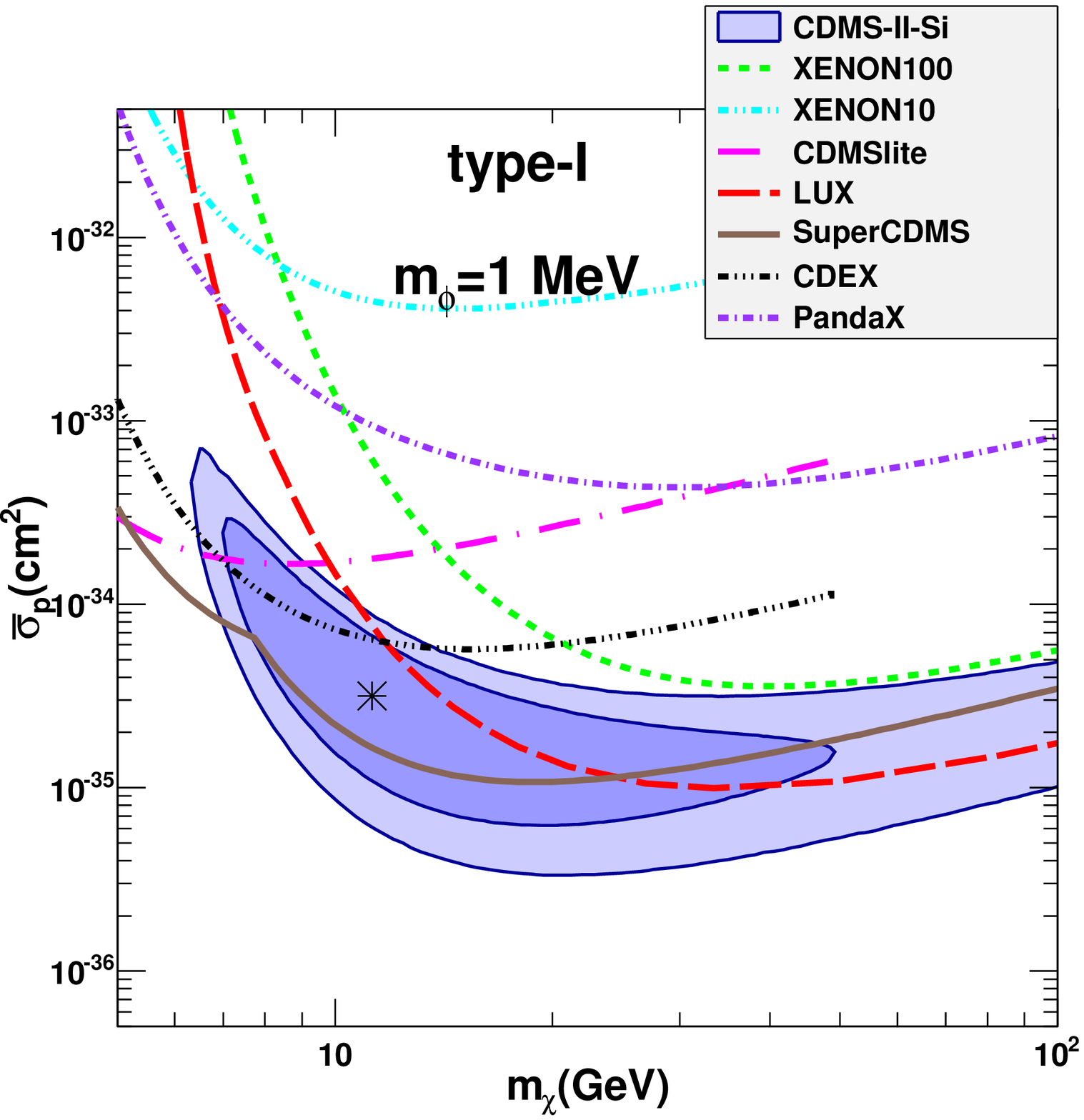}}

\subfloat[]{\includegraphics[scale=0.27]{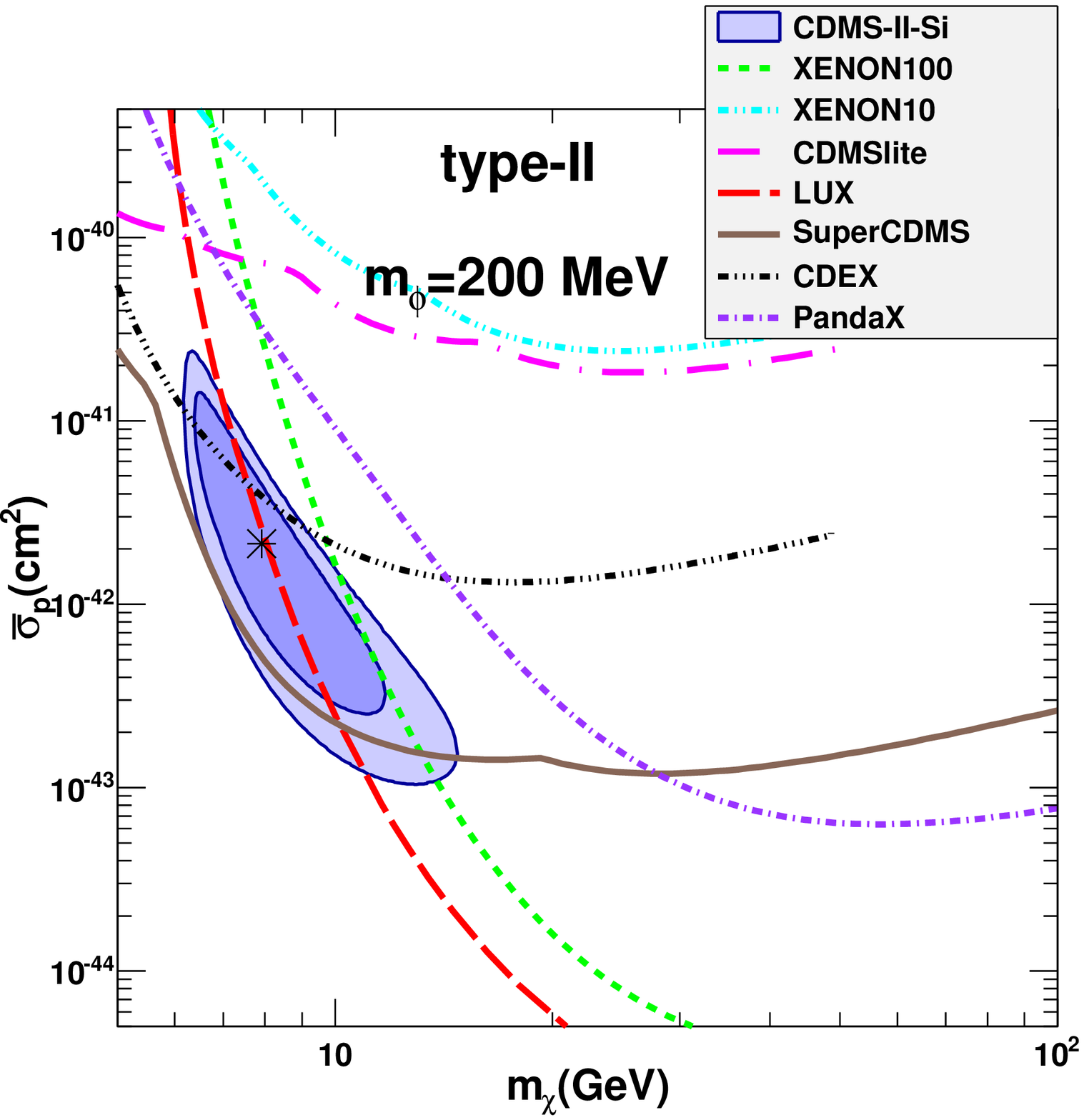}}
\subfloat[]{\includegraphics[scale=0.27]{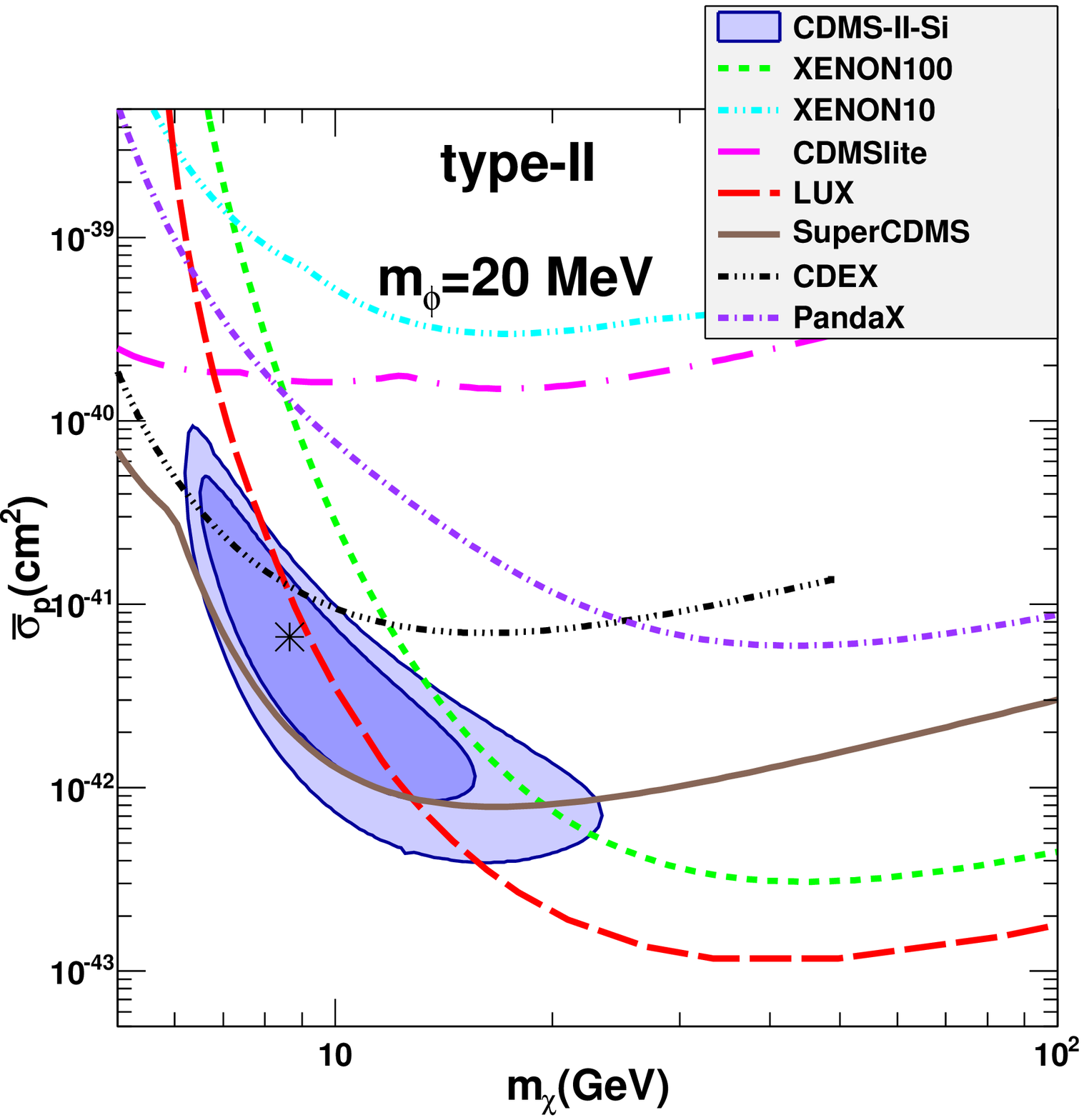}}
\subfloat[]{\includegraphics[scale=0.27]{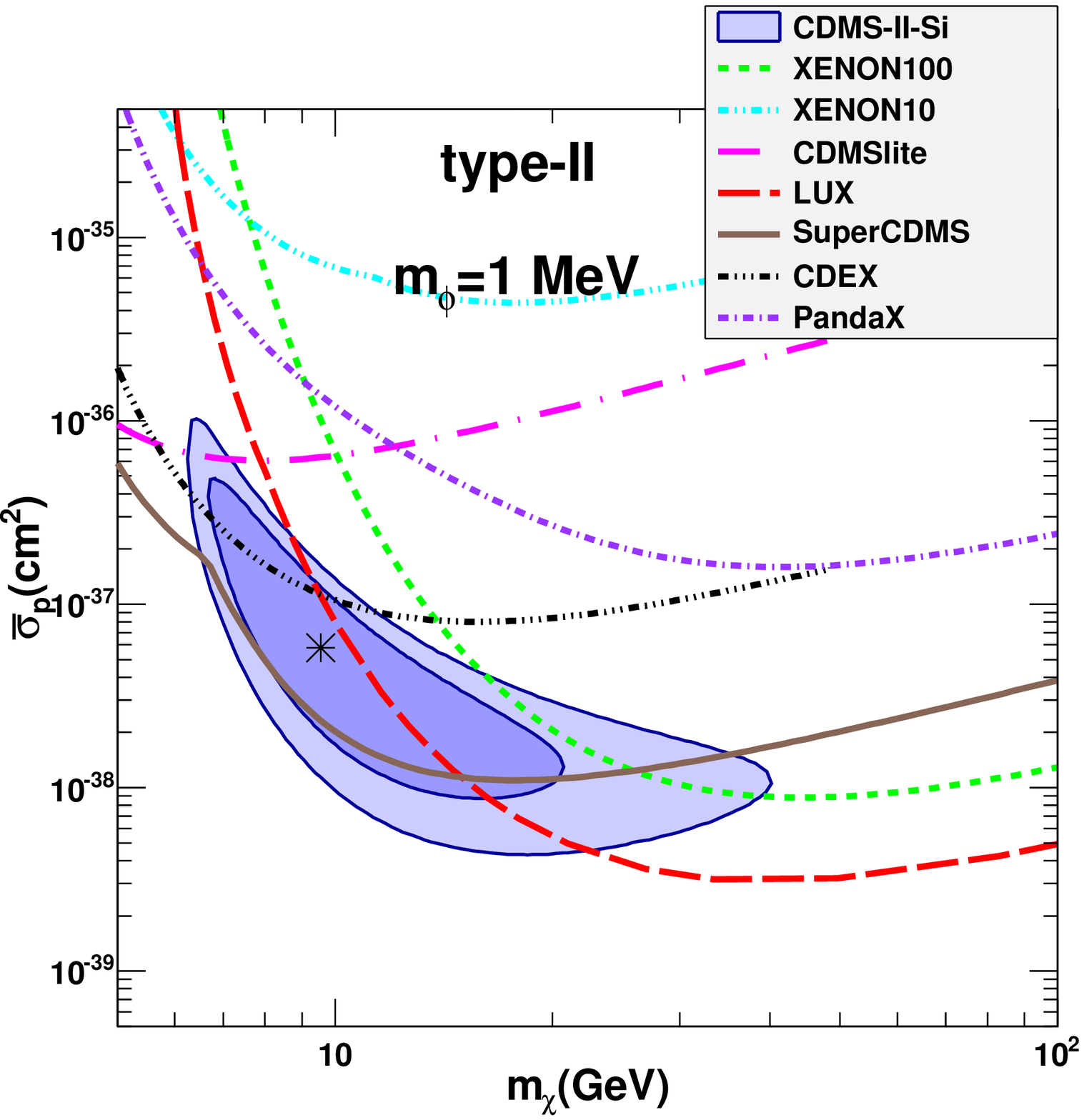}}

\subfloat[]{\includegraphics[scale=0.27]{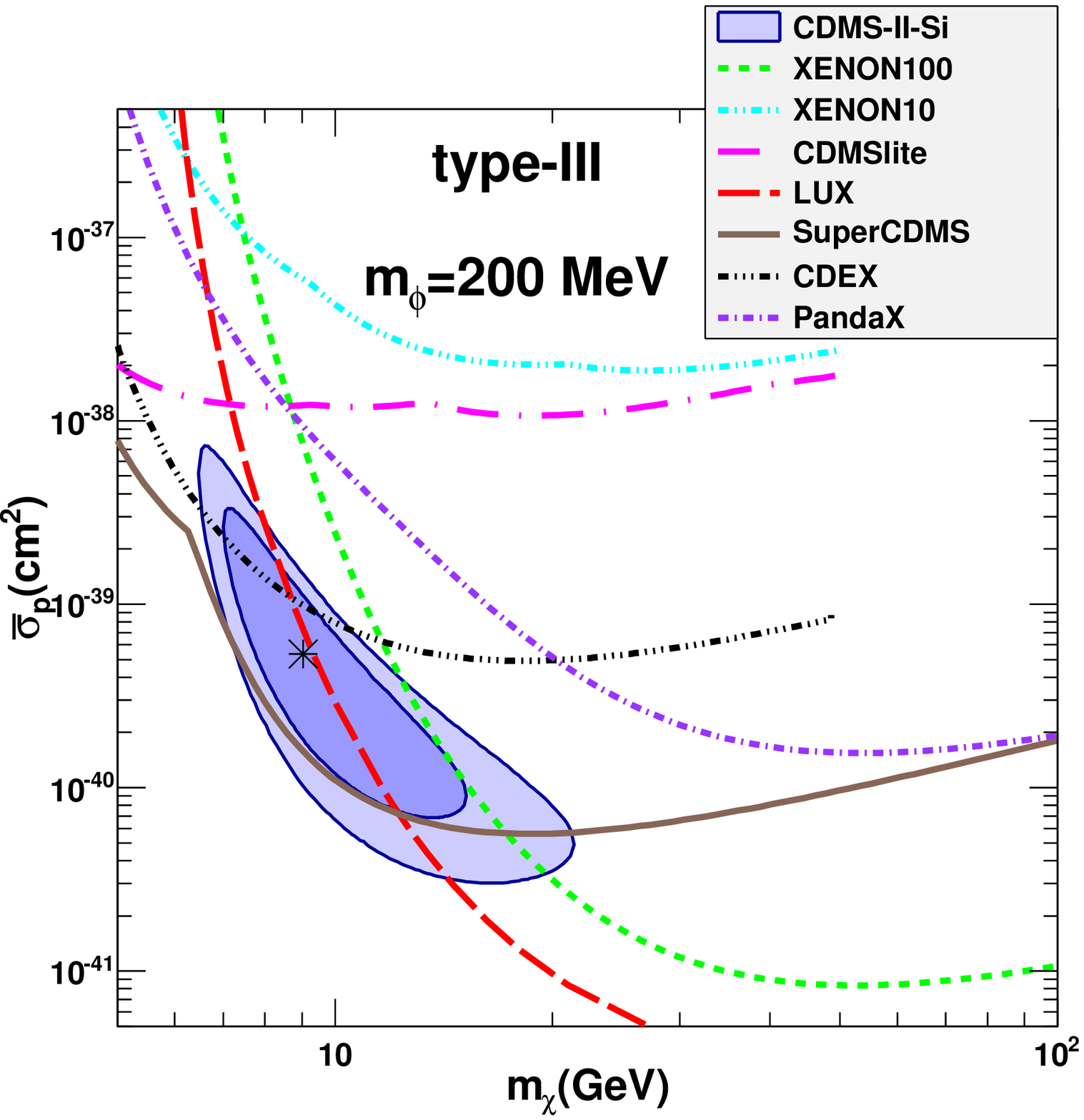}}
\subfloat[]{\includegraphics[scale=0.27]{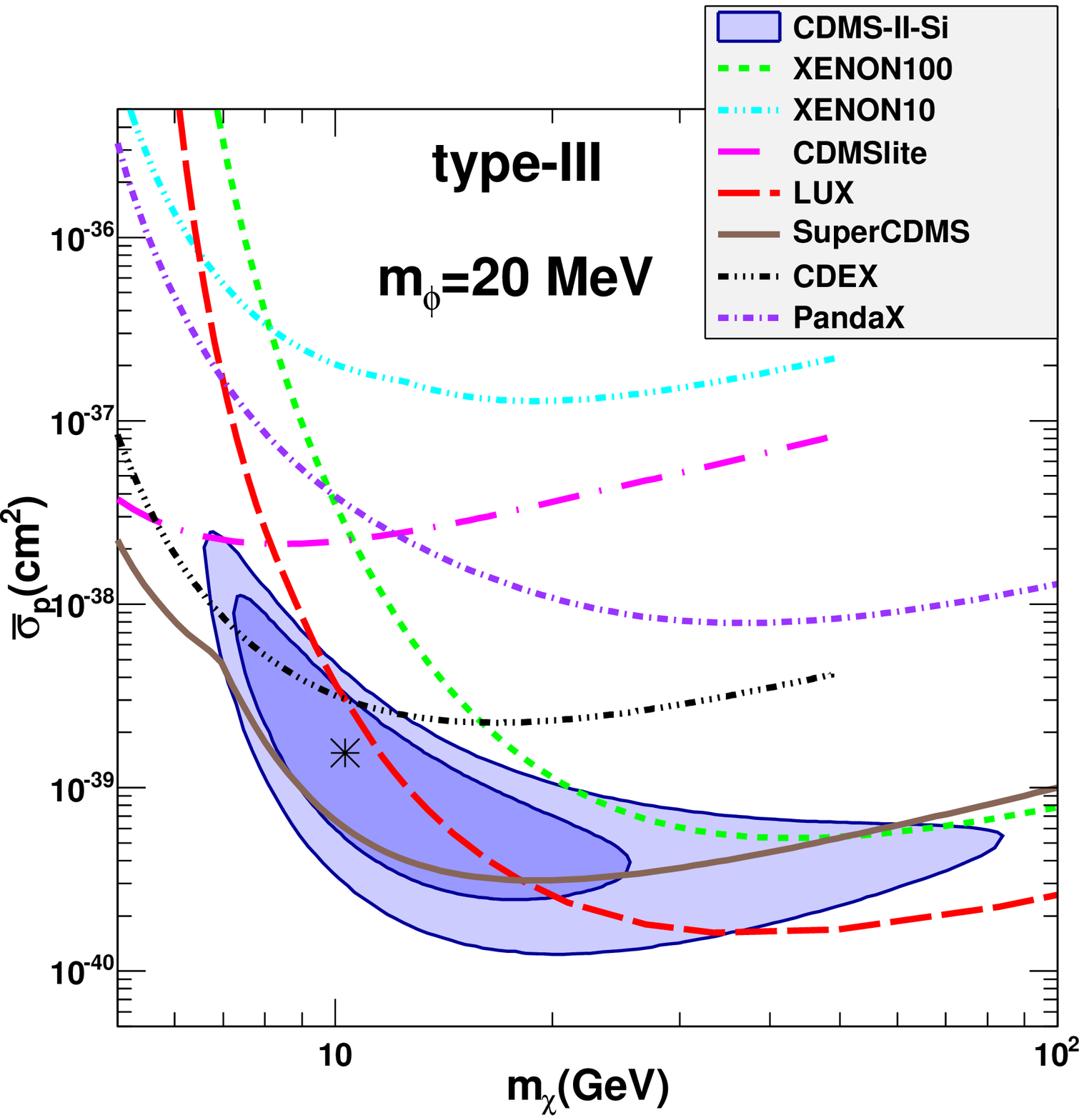}}
\subfloat[]{\includegraphics[scale=0.27]{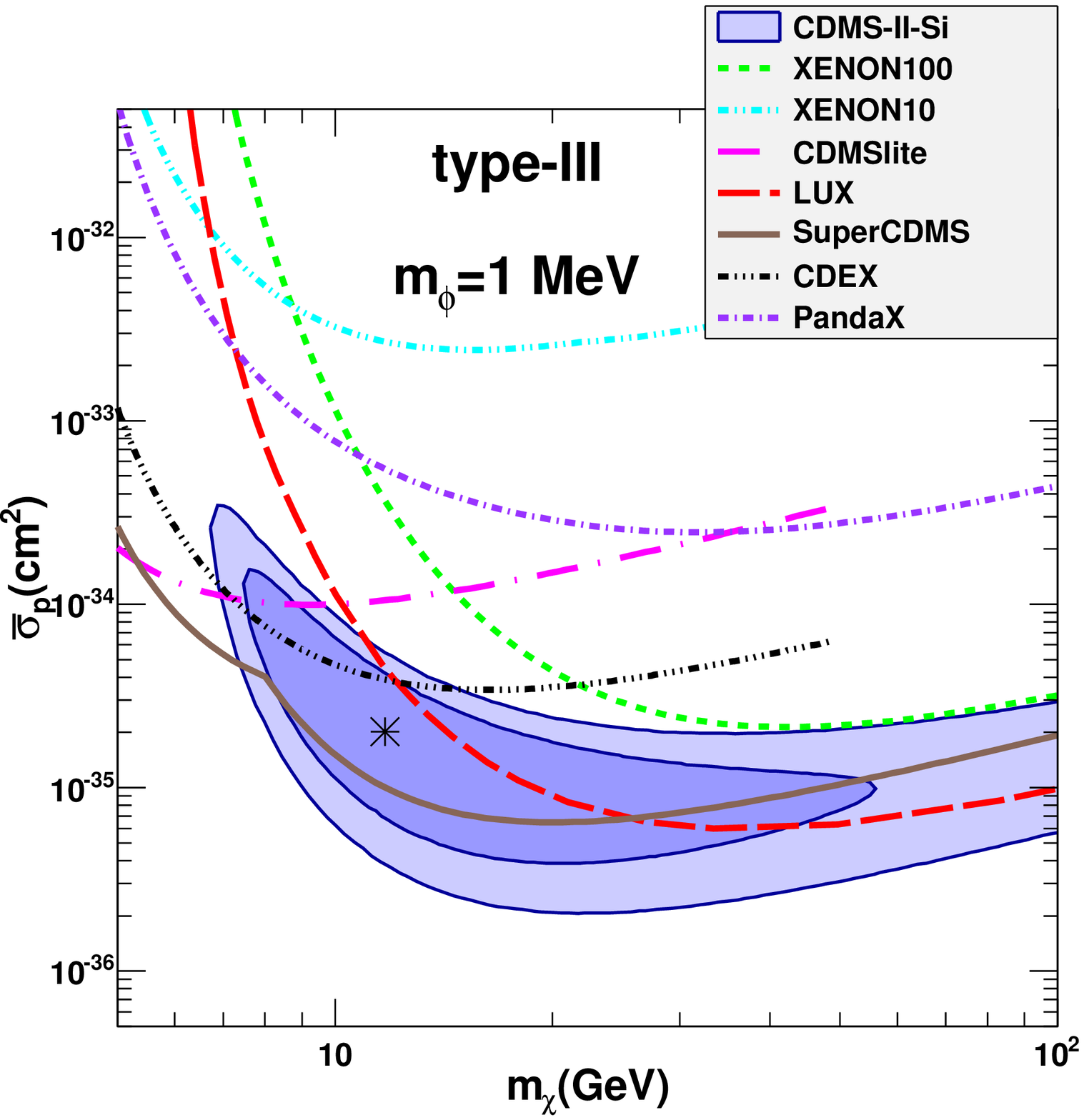}}
\caption{\label{fig:m_chi_sigmap}
The 68\% and 90\% CL favored regions from
CDMS-II-Si~\cite{Agnese:2013rvf}
as well as
90\% CL upper limits from
XENON100~\cite{Aprile:2012nq},
XENON10~\cite{Angle:2011th},
CDMSlite~\cite{Agnese:2013jaa},
LUX~\cite{Akerib:2013tjd},
SuperCDMS~\cite{Agnese:2014aze},
CDEX~\cite{Yue:2014qdu}
and
PandaX~\cite{Xiao:2014xyn}
in the ($m_{\chi},\sigma_p$) plane.
The best fit points of CDMS-II-Si are also shown as asterisks.
For type-I, II and III operators (from top to bottom) with
$m_{\phi}=$ 200, 20 and 1 MeV (from left to right).
The isospin violation parameter is fixed at $\xi=-0.7$.
}
\end{center}
\end{figure}
We first investigate  how the presence of a light mediator changes
the interpretation of  individual DM direct detection experiments,
with the focus on the results of  CDMS-II-Si.
We perform  a fit to
the CDMS-II-Si data using
the likelihood function of \eq{eq:EML} for
three typical cases of
very light ($m_\phi = 1$ MeV),
light ($m_\phi = 20$ MeV),
and
heavy ($m_\phi = 200$ MeV) mediators.
The case with ``very light '' mediator corresponds to the mediator mass
below the energy threshold of typical DM detection experiments,
and
the case with ``heavy''  mediator corresponds to a nearly contact DM-nucleus interaction.
The isospin-violation parameter  is fixed at $\xi=-0.70$ for future convenience,
as in this case the LUX constraint is maximally weakened.
Note that in the interpretation of  CDMS-II-Si data alone,
a different choice of $\xi$ only leads to a rescaling of $\bar\sigma_{p}$.
In Fig.~\ref{fig:m_chi_sigmap},
the best-fit values of $m_{\chi}$ and $\bar\sigma_{p}$
and
the regions favored by the  data in
the ($m_{\chi},\bar\sigma_p$) plane at
$68\%$ and $90\%$ CL  are shown  for
the operators of type-I, -II and -III.
The effect of  light mediators  on
the $90\%$~CL upper limits  of the experiments
such as
SuperCDMS,  CDMSlite, CDEX, LUX, XENON100/10 and  PandaX
are also calculated and  shown in Fig.~\ref{fig:m_chi_sigmap}.
The upper limits from SuperCDMS is calculated using the maximum gap method.
As expected,
the  value of $m_{\chi}$ favored by the CDMS-II-Si data
increases when the mediator becomes lighter,
and the upper limits from other experiments becomes
weaker towards high DM particle mass.

For type-I operators,
the favored value of $m_{\chi}$ changes dramatically with a decreasing $m_\phi$.
At $68\%$ CL,
for a heavy mediator $m_\phi = 200~\hbox{MeV}$,
the CDMS-II-Si favored DM particle mass is in the range $\sim 7-15$ GeV.
For a smaller $m_\phi = 20~\hbox{MeV}$,
the allowed value of $m_{\chi}$ can reach  $\sim 23$ GeV.
For a tiny  $m_\phi = 1~\hbox{MeV}$,
it can reach   $\sim 50$ GeV.
At $90\%$ CL,
the effect of light mediator is even more significant.
For $m_\phi = 20~\hbox{MeV}$,
the allowed value of $m_{\chi}$ is already larger than $70$ GeV.
For $m_\phi = 1~\hbox{MeV}$,
the value of $m_{\chi}$ can be  larger than $100$ GeV.
As it can be seen from \fig{fig:m_chi_sigmap},
the upper limits from other experiments also
change  significantly with different choices of $m_{\phi}$.
In general,
when the mediator is lighter,
the shape of the  exclusion curve  becomes more flat toward large $m_{\chi}$ region,
which indicates that the limits are  relatively weaker compared with
that at low $m_{\chi}$ region.
This effect of light mediator can be clearly seen in the exclusion limits of  LUX.
For type-I operators, when $m_{\phi}=200$ MeV,
the LUX limit at $m_{\chi}=10~(50)$ GeV is
$\sim 3\times 10^{-40}~(5\times 10^{-42})~\text{cm}^{2}$,
which indicates that
the limit at $m_{\chi}=50$ GeV is lower than that at 10 GeV by a factor of $\sim60$.
When  $m_{\phi}=20$ MeV, this factor is reduced to $\sim17$,
and for  $m_{\phi}=1$ MeV, this factor is further reduced to $\sim10$.
Similar observations can be obtained for other experiments.
Although
only a small portion of the CDMS-II-Si favoured region
is allowed after the  constraints from  LUX and SuperCDMS,
the allowed region is enlarged when $m_{\phi}$ is smaller,
which shows the possibility that
the presence of a light mediator can
relax the tension between these experimental results.

For type-II and type-III operators,
the conclusions are similar.
For these type of operators,
light mediators also enlarge the allowed range for the DM particle mass.
For type-II operators,
the upper limits on the DM particle mass are
$\sim 12 (17)\GeV$ at 68\% (90\%) CL for $m_{\phi} = 200$ MeV,
and
the upper limits change to $\sim 20 (40)\GeV$ at 68\% (90\%) CL  for $m_{\phi} = 1 \MeV$.
Compared with type-I operators,
the allowed DM particle mass is smaller,
which is in agreement with the analytical analysis in section \ref{sec:mediator-mass}.
For type-III operators,
the obtained upper limits are quite similar to the case with type-I operators,
which can be understood from Eqs. (\ref{eq:typerelation})-(\ref{eq:typerelation2}).
The upper limits on the DM particle mass are
$\sim 16~(22)\GeV$ at 68\% (90\%) CL  for $m_{\phi} = 200$,
and the upper limits  are $\sim 60~(100)\GeV$ at 68\%~(90\%) CL for $m_{\phi} = 1 \MeV$.

\begin{figure}\begin{center}
\subfloat[]{\includegraphics[scale=0.27]{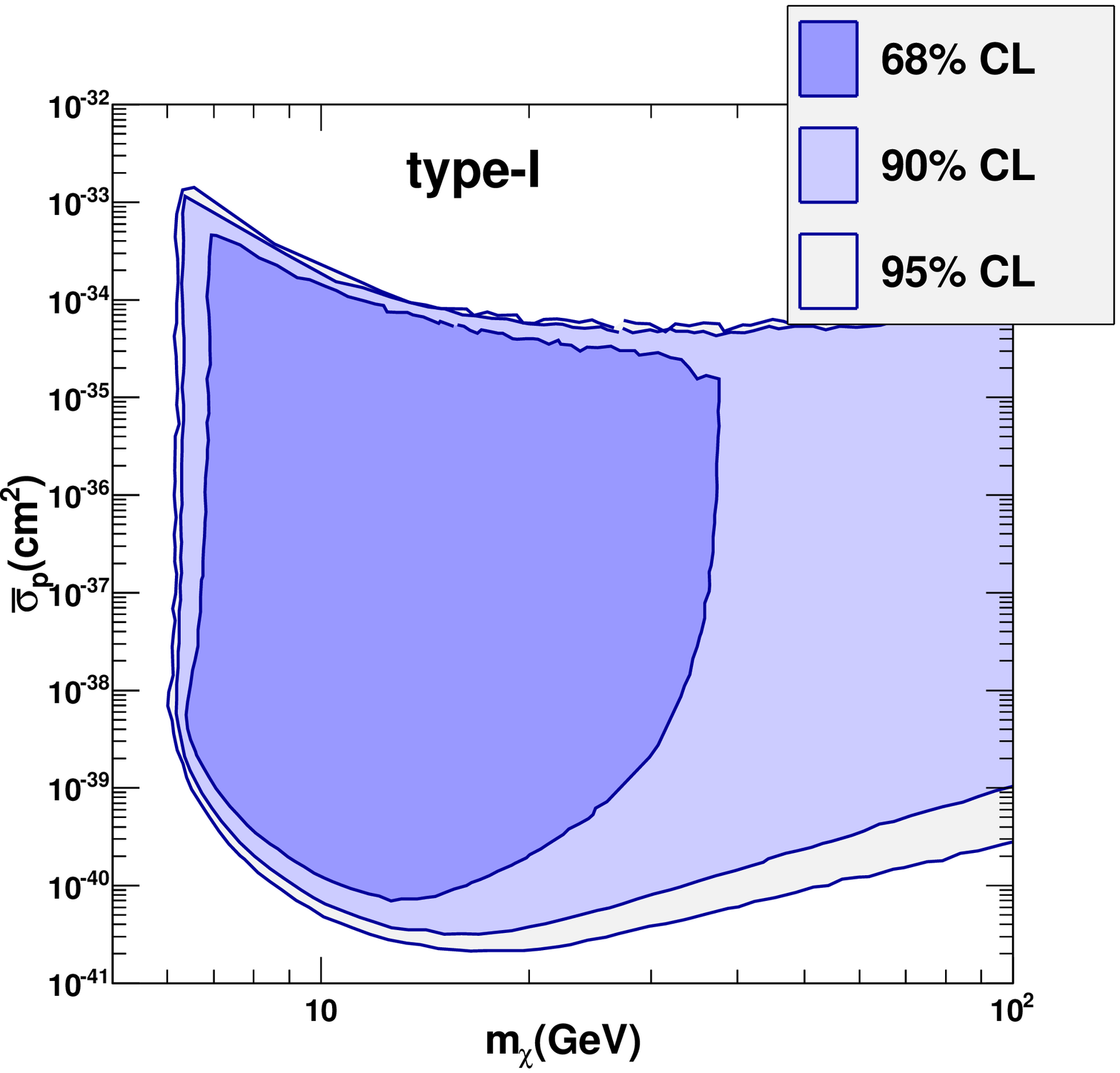}}
\subfloat[]{\includegraphics[scale=0.27]{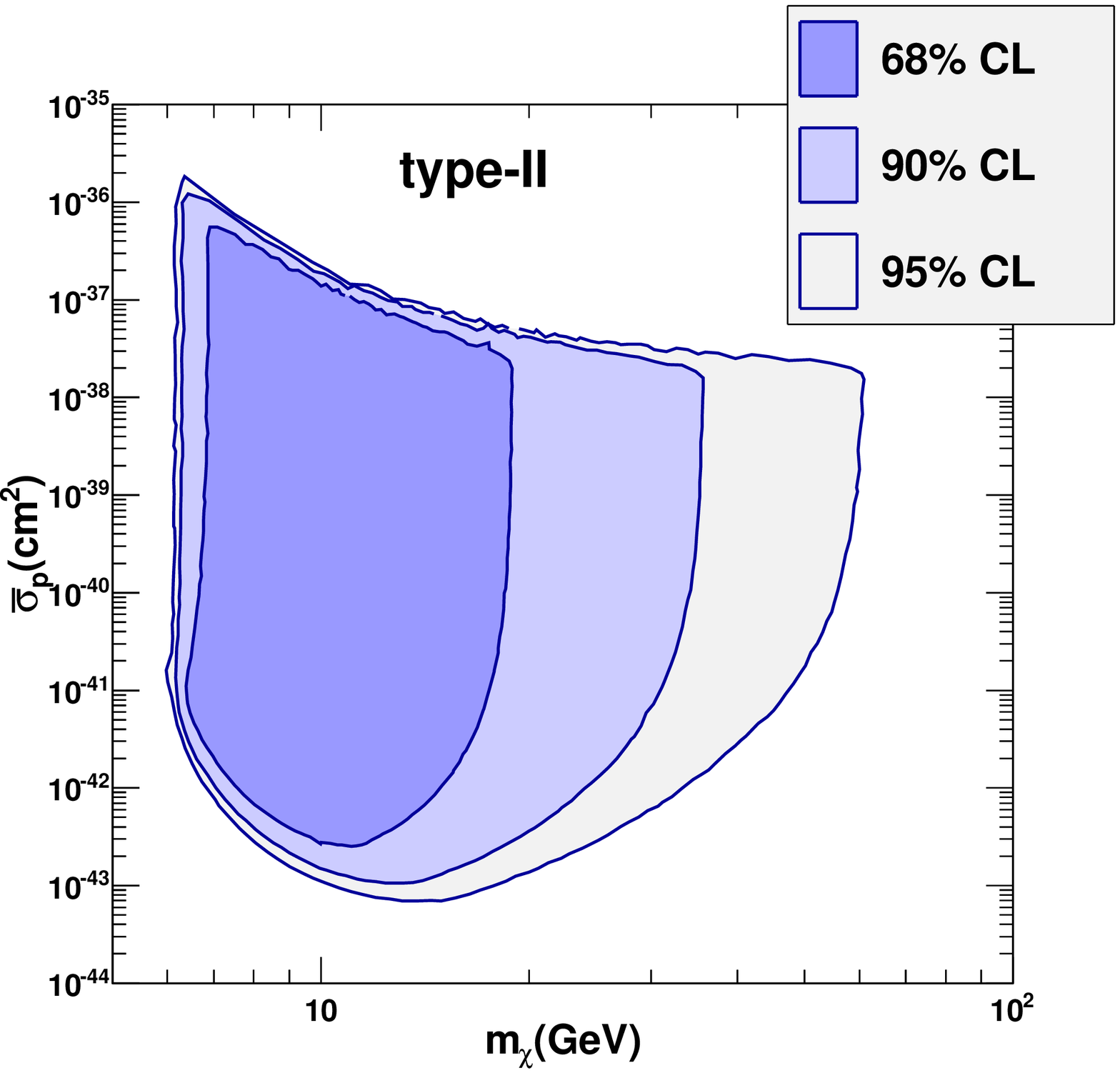}}
\subfloat[]{\includegraphics[scale=0.27]{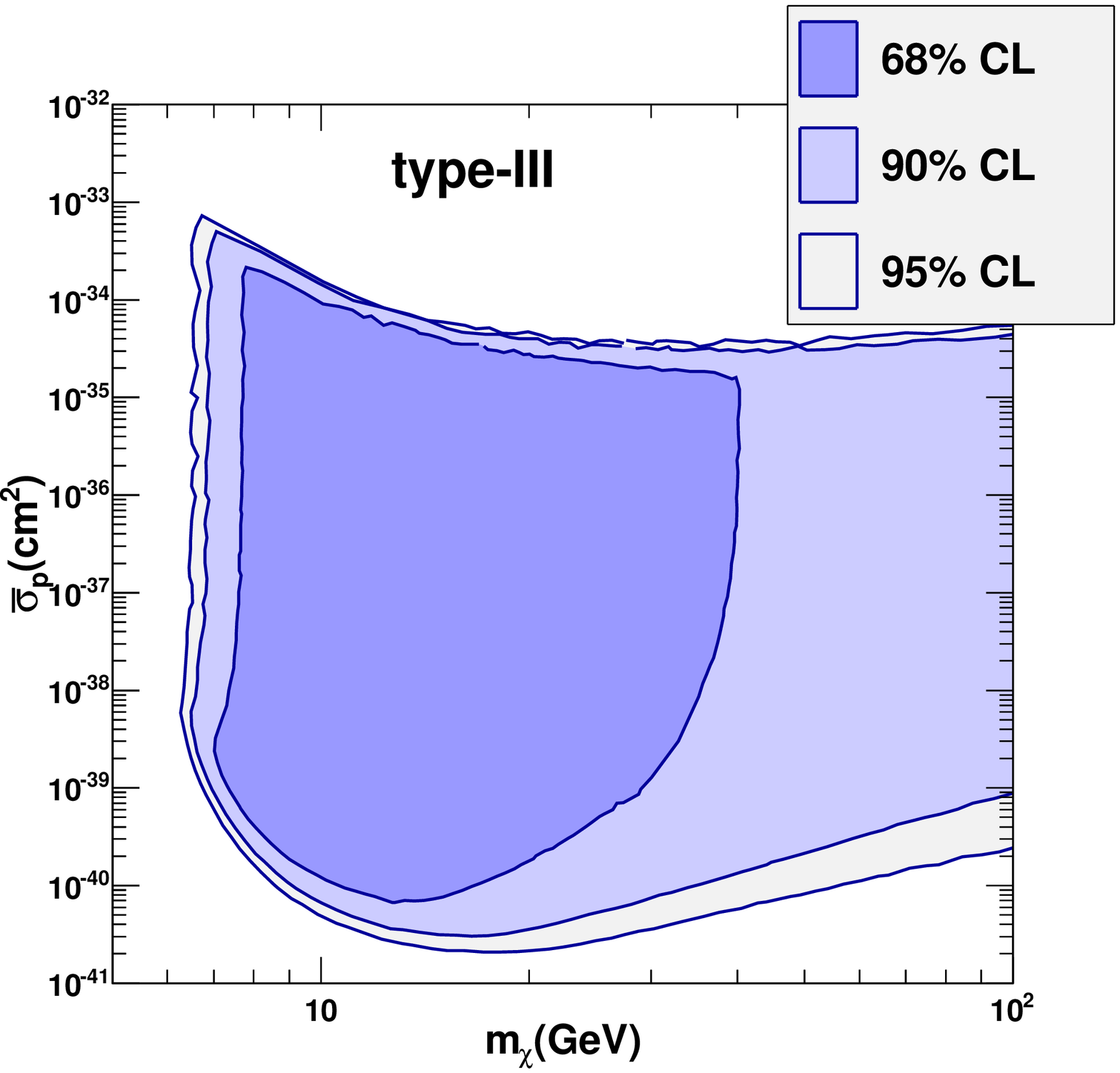}}

\subfloat[]{\includegraphics[scale=0.27]{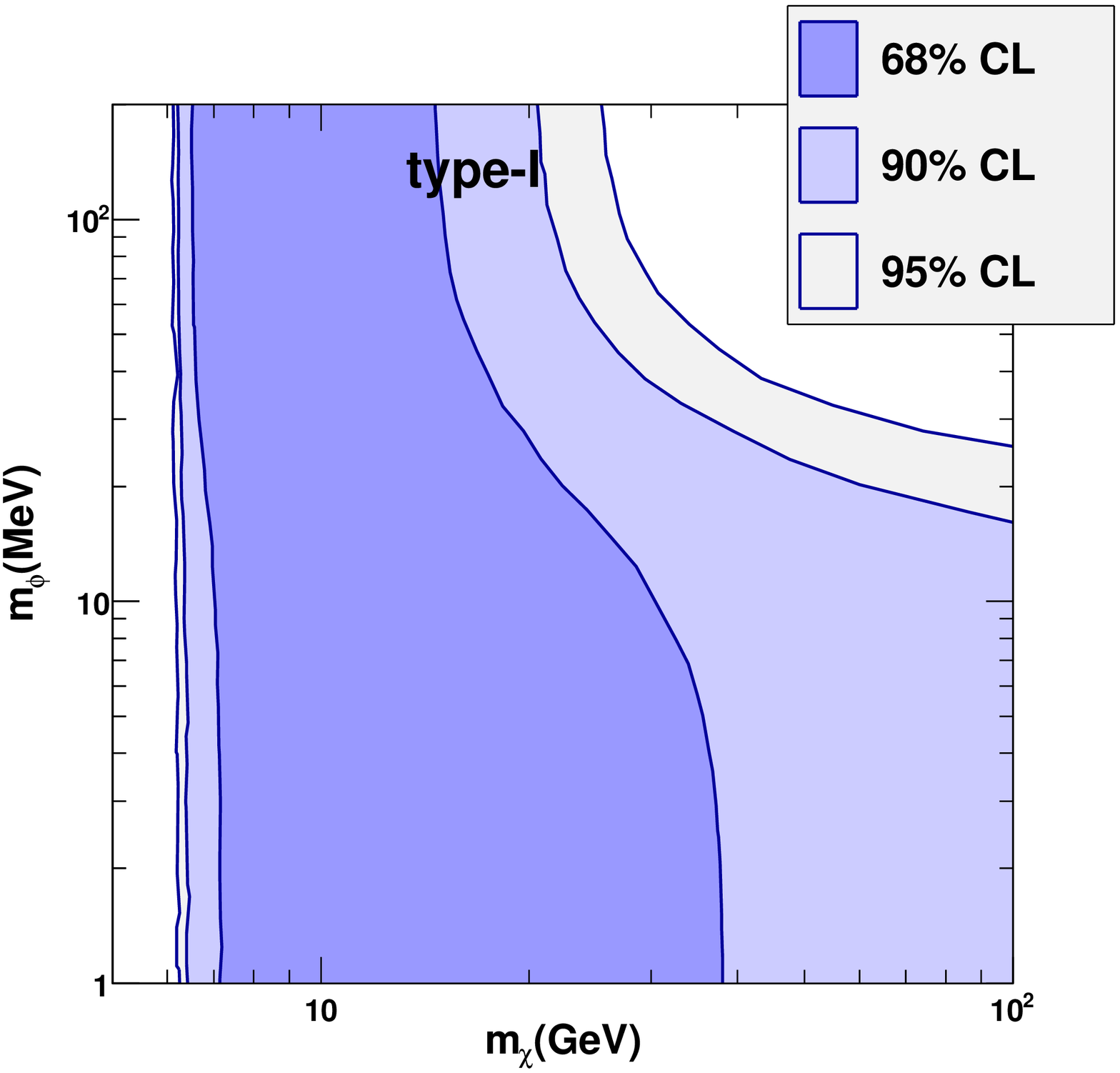}}
\subfloat[]{\includegraphics[scale=0.27]{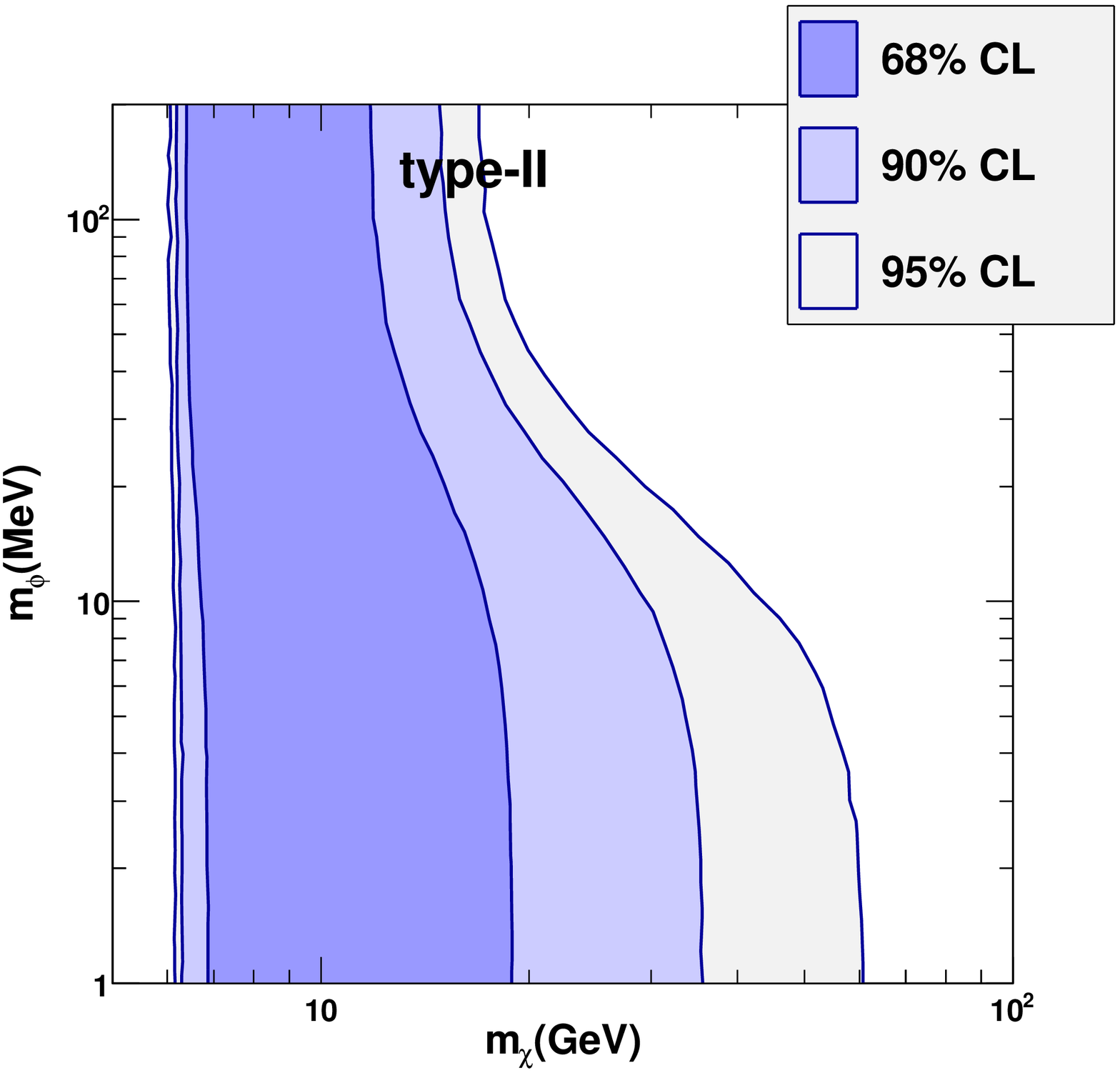}}
\subfloat[]{\includegraphics[scale=0.27]{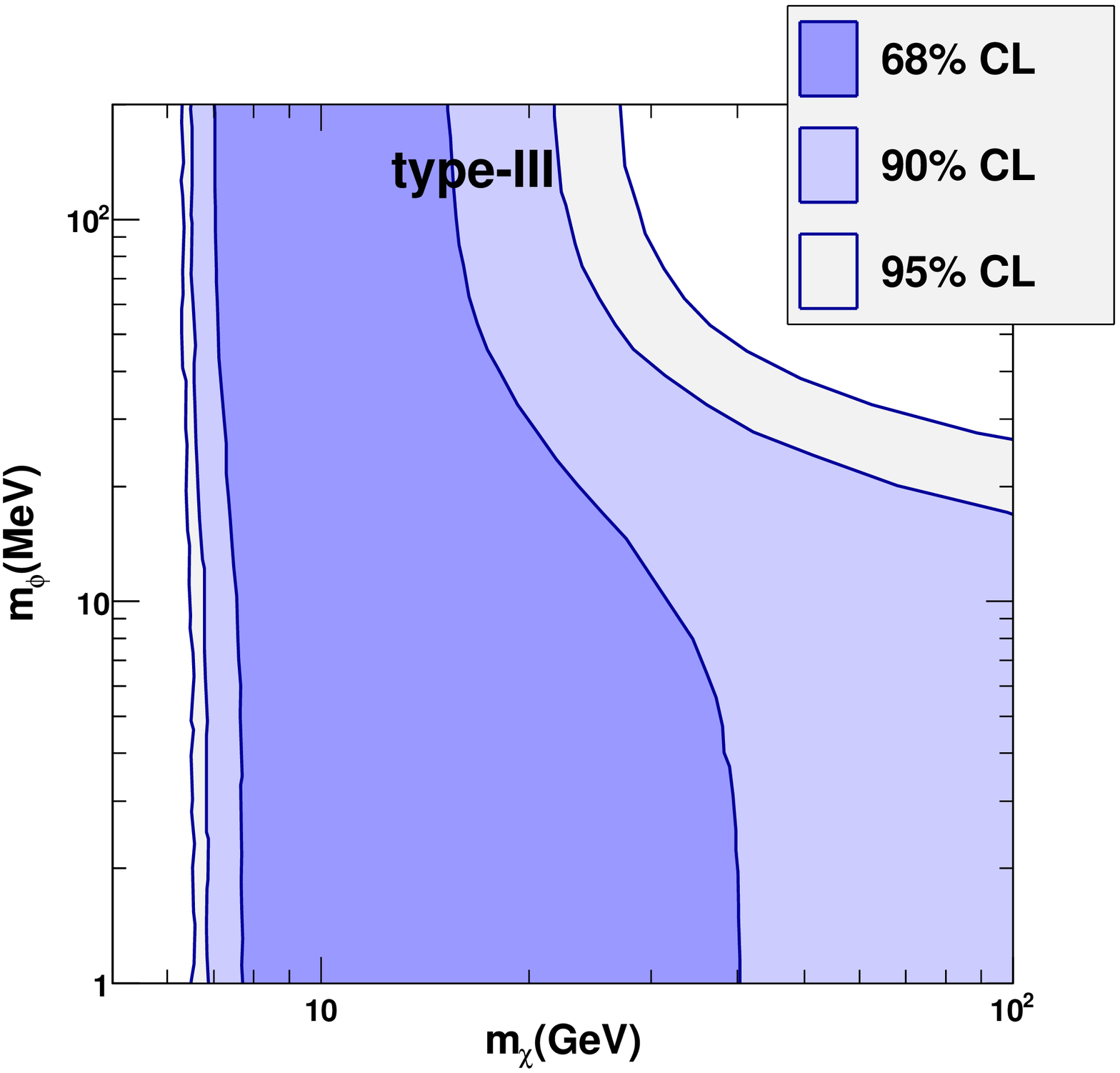}}
\end{center}
\caption{\label{fig:CDMSSi}
The regions of parameter space allowed by the CDMS-II-Si data
at 68\%, 90\% and 95\% CLs in
($m_{\chi},\sigma_p$) plane (top panels) and
($m_{\chi},m_\phi$) plane (bottom panels)
for three type of operators.
The isospin-violation parameter is
is fixed at $\xi= -0.7$.
}
\end{figure}

In the next step,
we allow the value of $m_{\phi}$ to be free
(with a cut off at $m_{\phi}\geq$ 0.2 MeV) and
to be determined by fitting the CDMS-II-Si data.
The allowed regions of parameter space
are shown
in $(m_{\chi},m_{\phi})$ plane and ($m_{\chi},\bar\sigma_{p}$) plane, respectively,
in \fig{fig:CDMSSi}.
For all the operators under consideration,
the value of $m_{\chi}$ increases with
a decreasing $m_{\phi}$.
However, at $m_{\phi}\approx 5$ MeV, the increase saturates,
corresponding to the case B in \eq{eq:Sexp}.
At $68\%$ CL,
the maximally allowed DM particle masses are
$\sim 40$ GeV, $\sim 20$ GeV and $\sim 40$ GeV
for type-I, type-II and type-III  operators, respectively.
Note that at higher confidence levels such as $95\%$ CL,
the allowed value of $m_{\phi}$ can exceed 100 GeV for type-I and type-III operators.
Since a small $m_{\phi}$ can greatly enhance the scattering amplitude,
the allowed range of $\bar\sigma_{p}$ can vary in a large range of a few order of magnitudes.
This situation is also shown in \fig{fig:CDMSSi}.

\begin{figure}
\begin{center}
\subfloat[]{\includegraphics[scale=0.27]{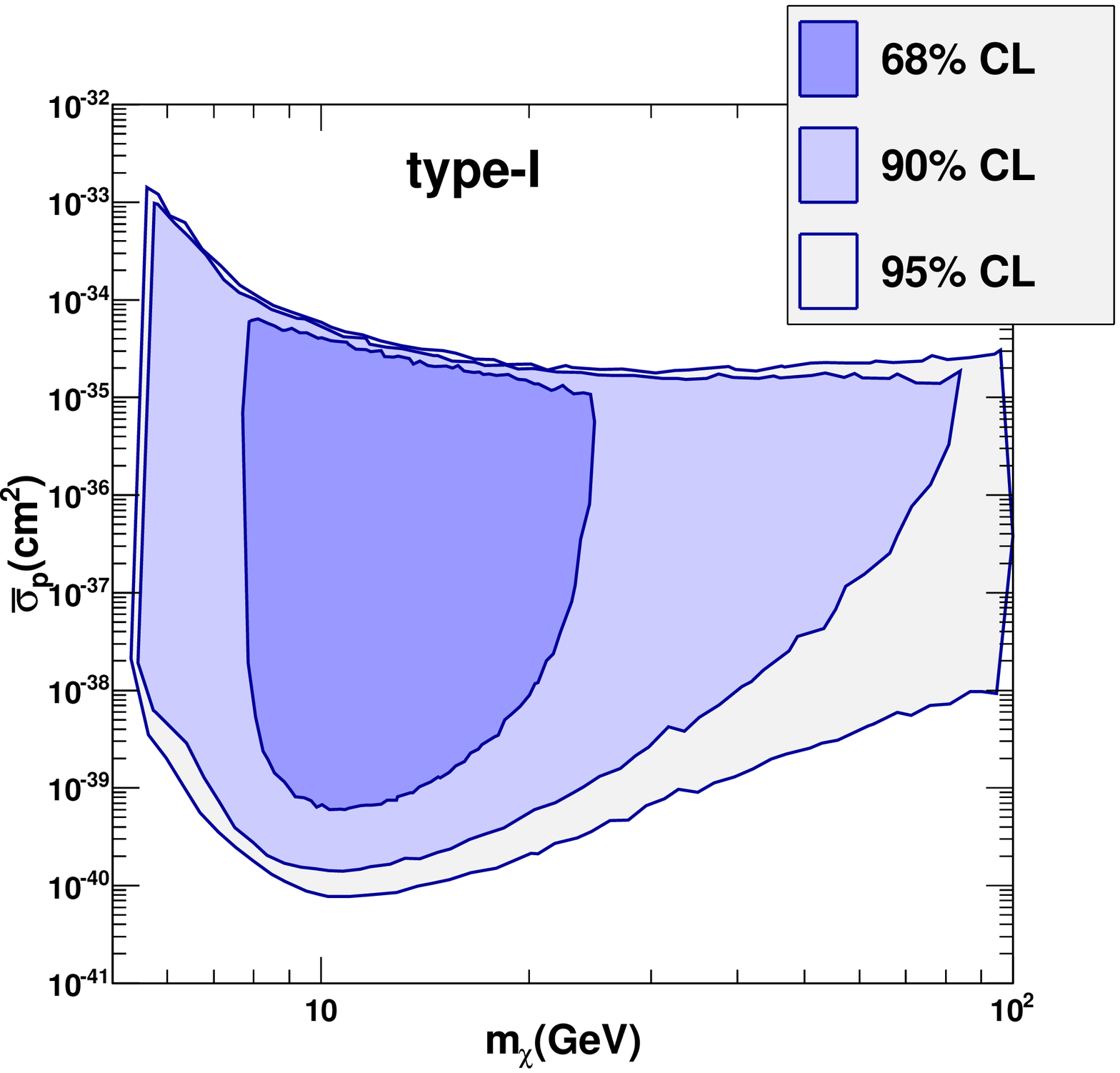}}
\subfloat[]{\includegraphics[scale=0.27]{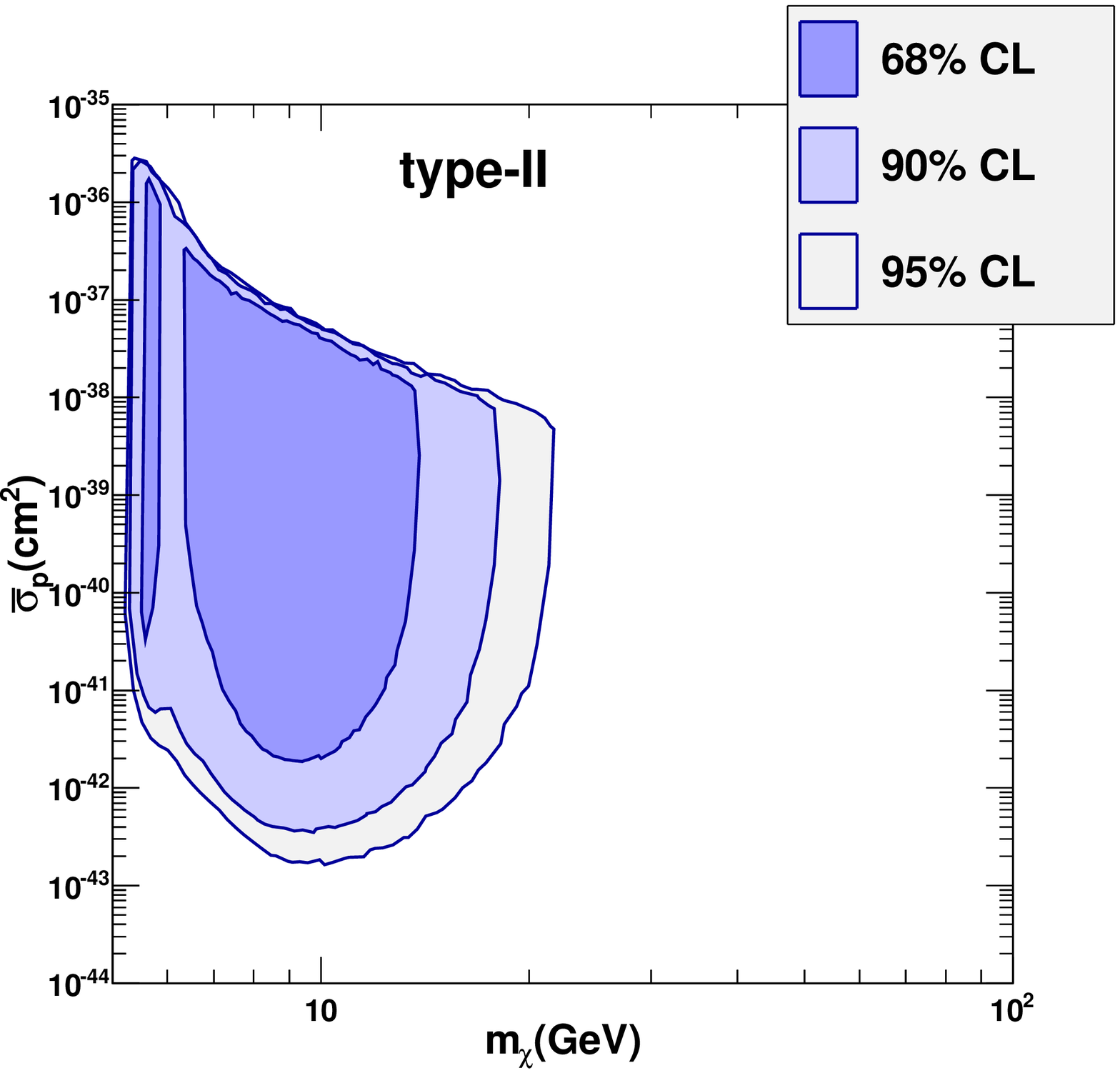}}
\subfloat[]{\includegraphics[scale=0.27]{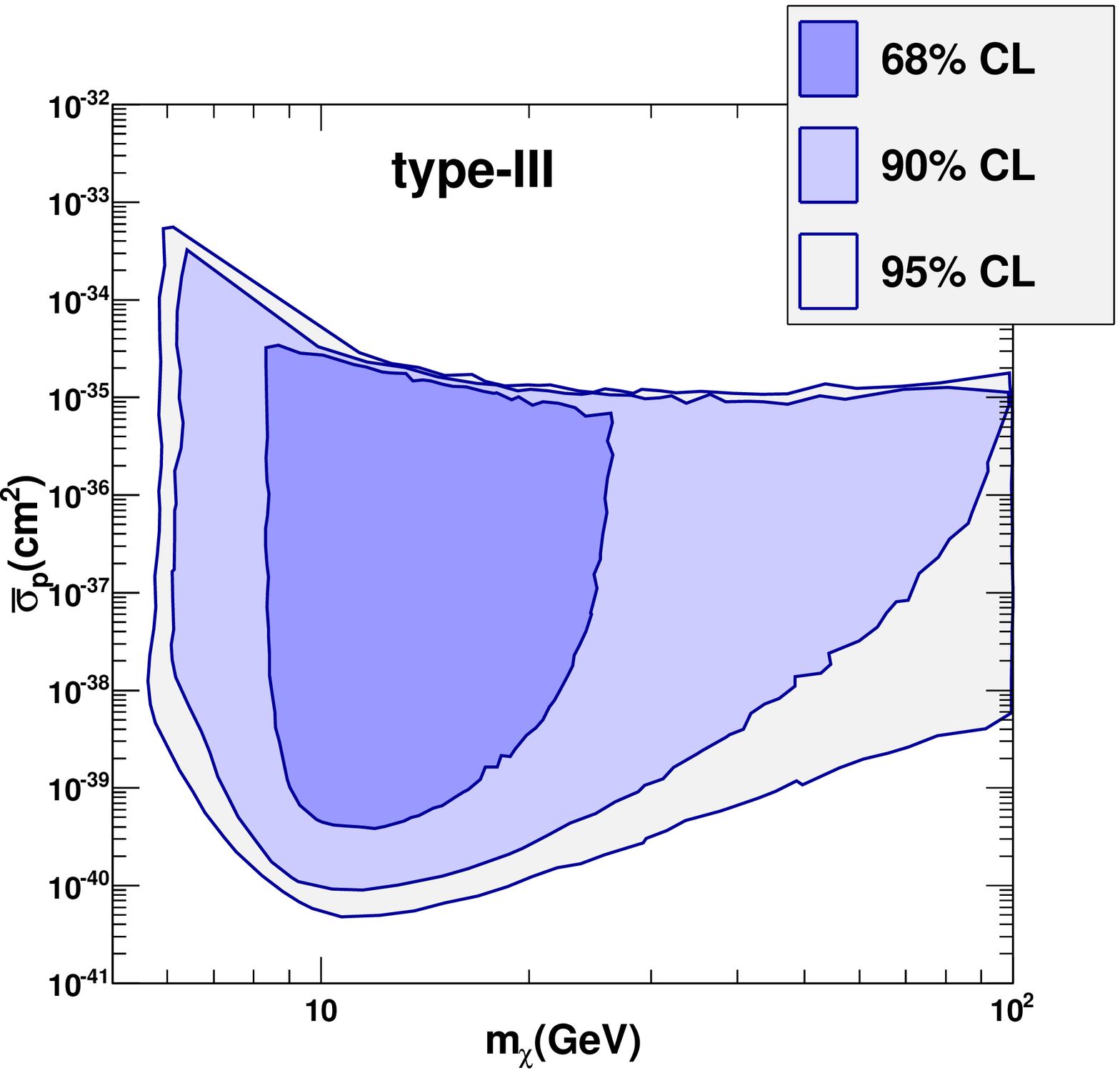}}

\subfloat[]{\includegraphics[scale=0.27]{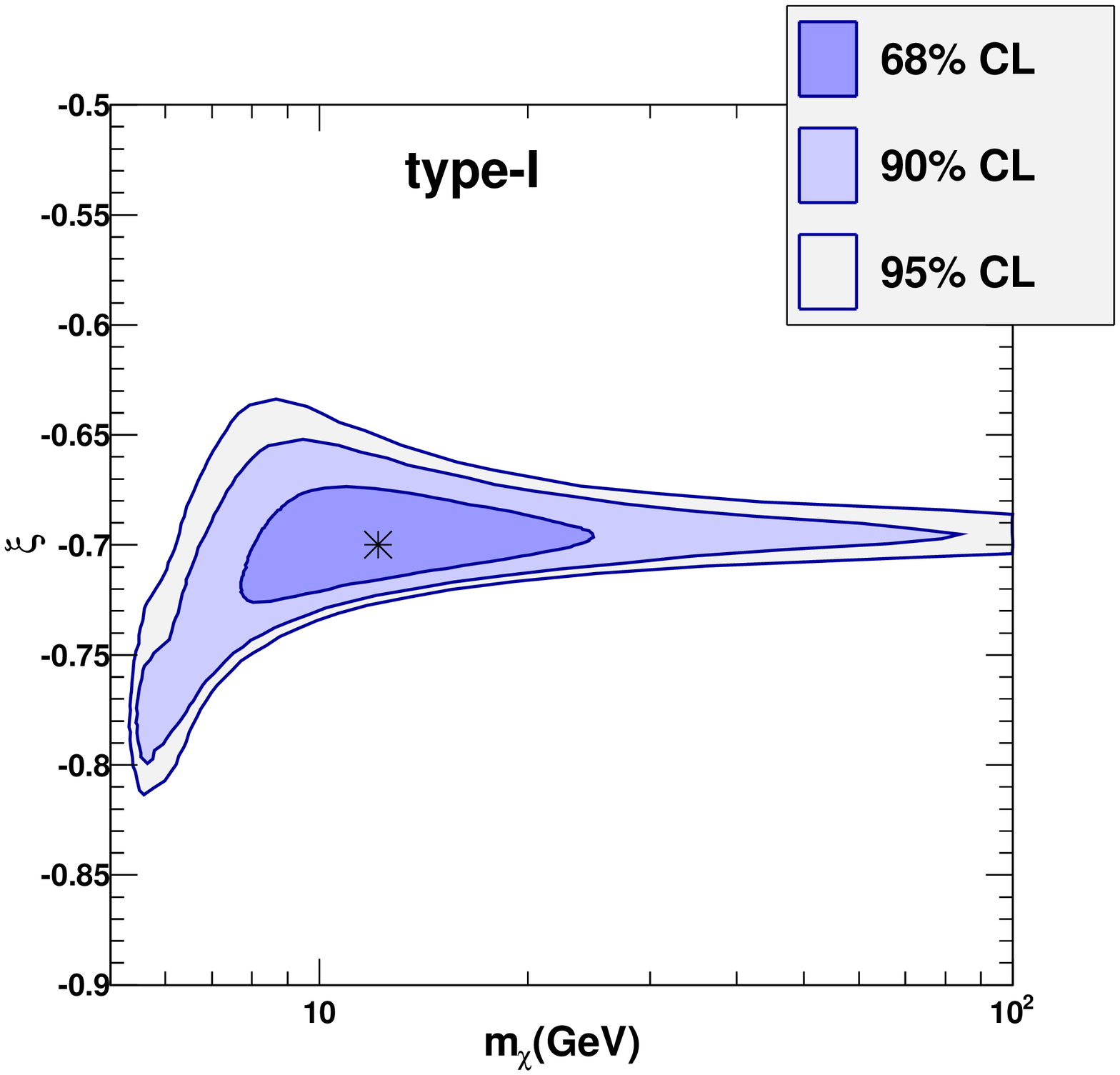}}
\subfloat[]{\includegraphics[scale=0.27]{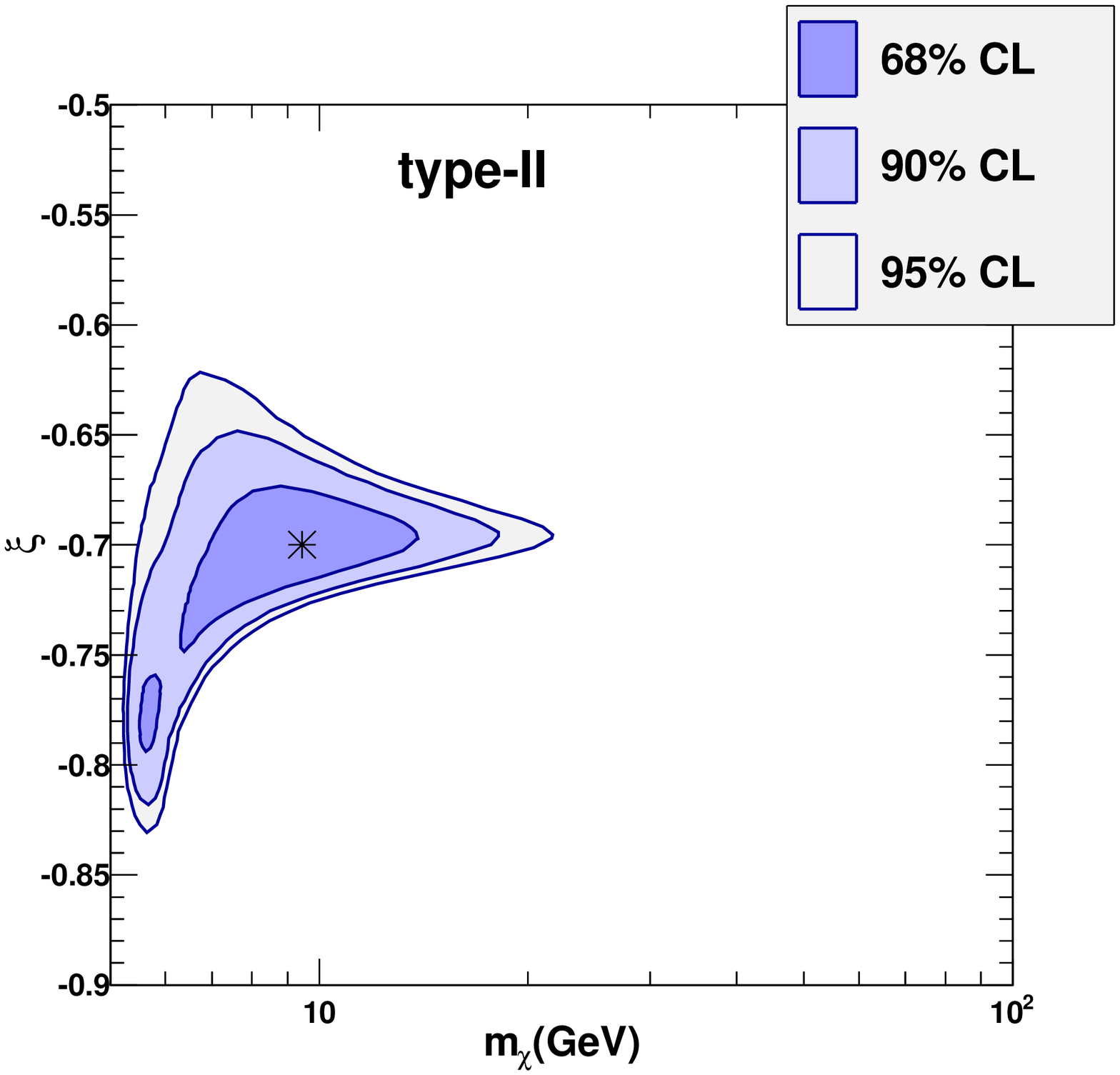}}
\subfloat[]{\includegraphics[scale=0.27]{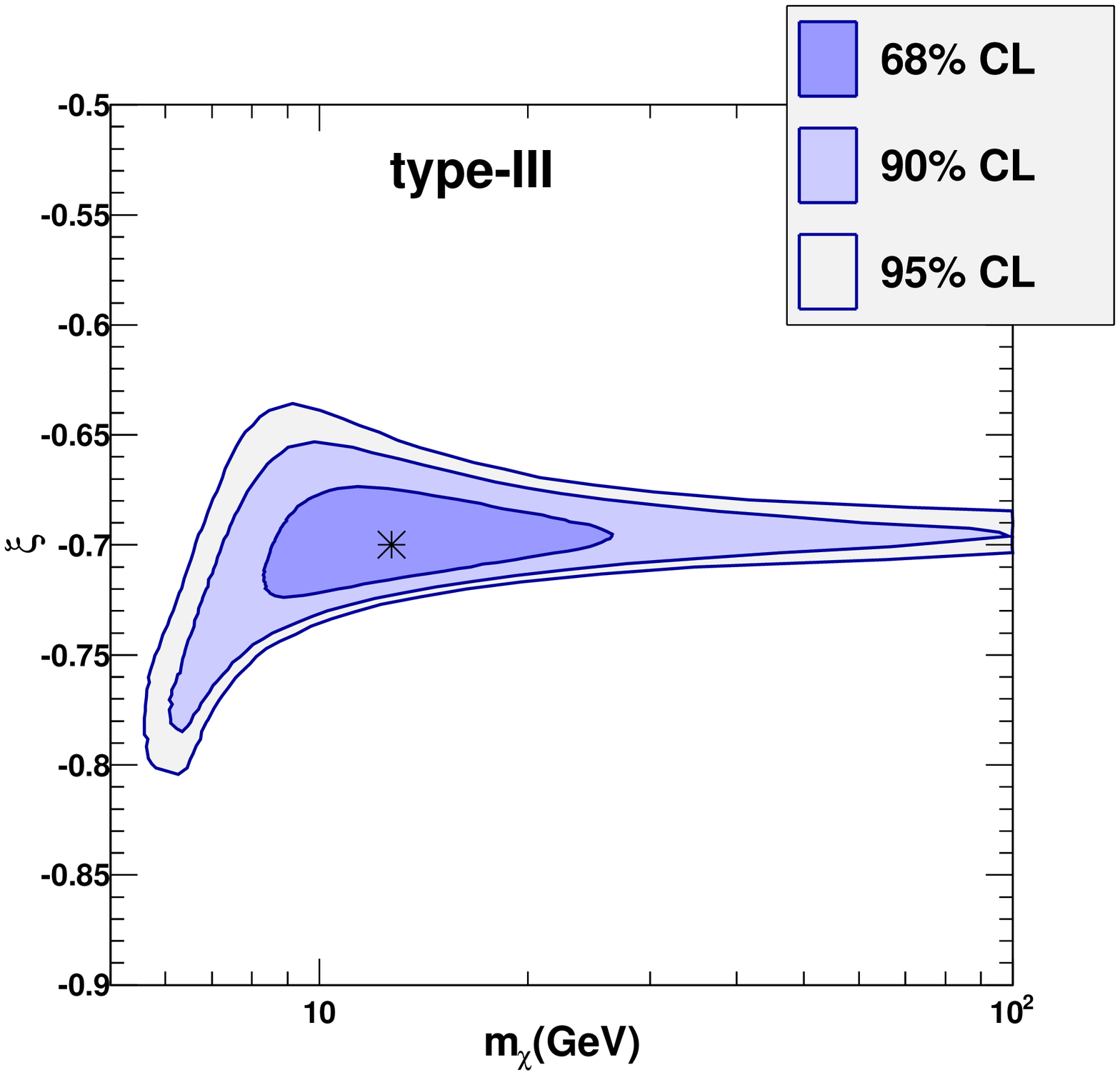}}

\subfloat[]{\includegraphics[scale=0.27]{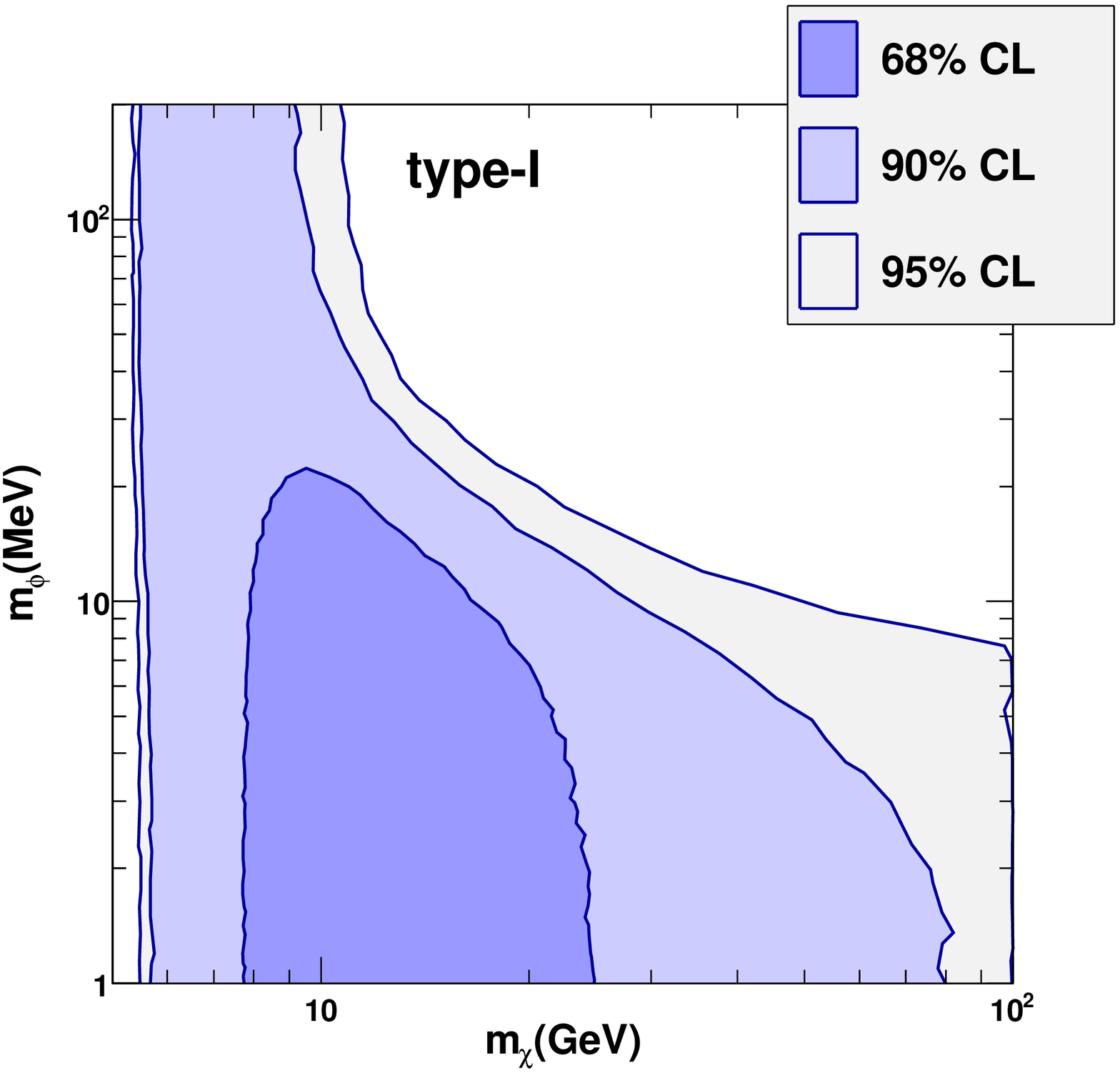}}
\subfloat[]{\includegraphics[scale=0.27]{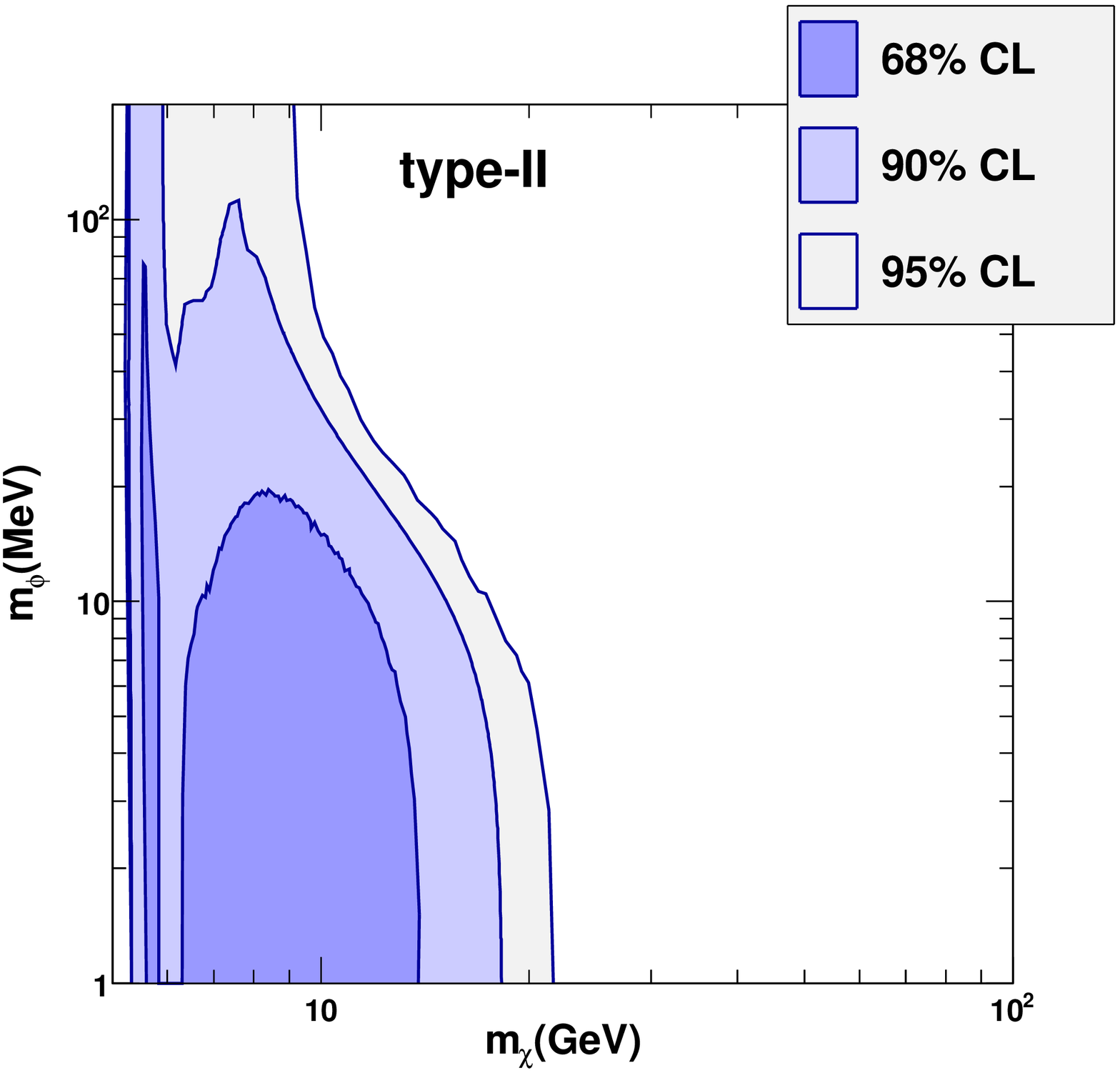}}
\subfloat[]{\includegraphics[scale=0.27]{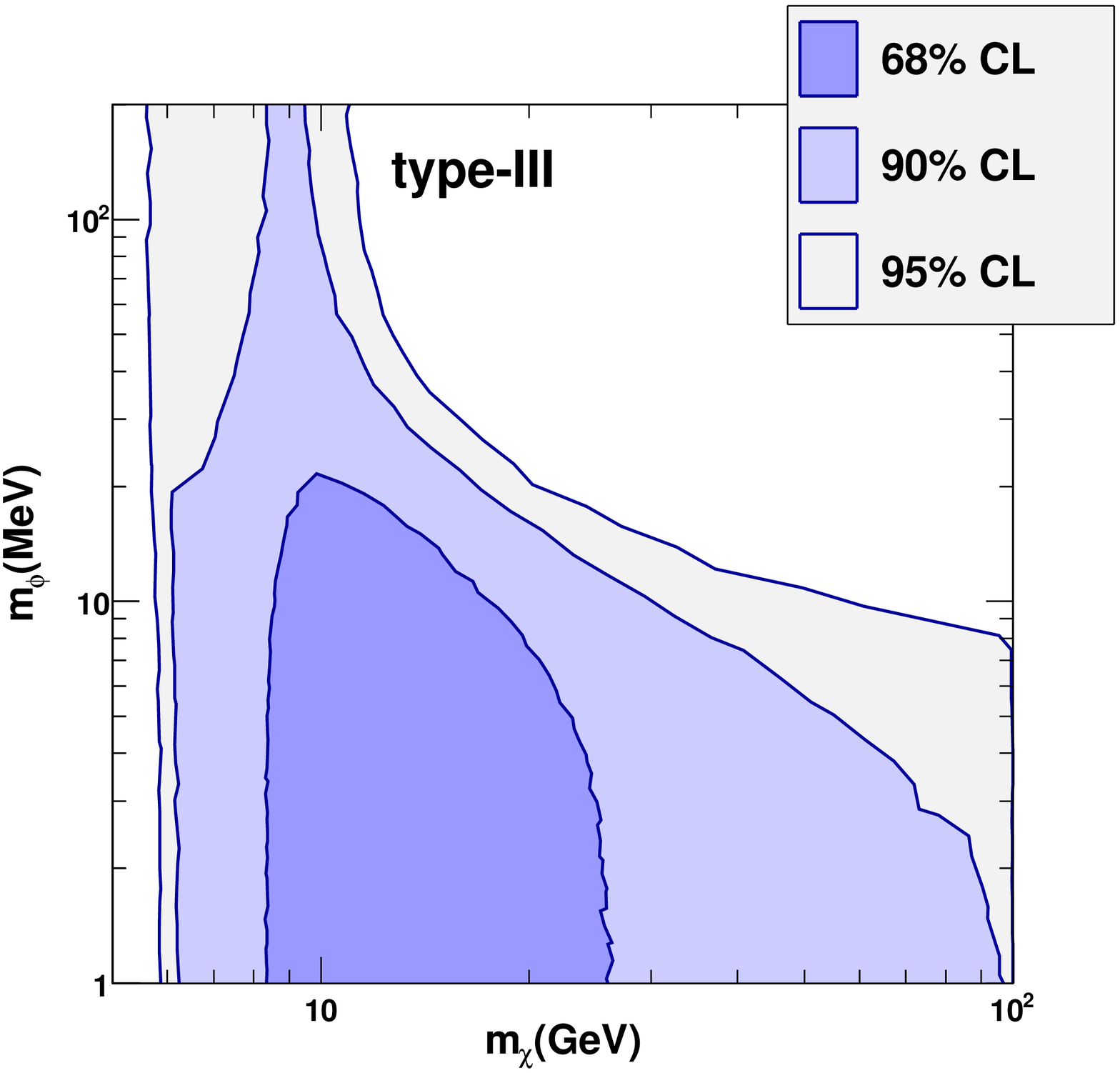}}
\end{center}
\caption{\label{fig:global}
The favored regions (68\%, 90\% and 95\% CL) from the global fits on the combination of
CDMS-II-Si,
LUX and
SuperCDMS for three type of opeartors.
(Top panels) the favored regions  in the $(m_{\chi},\sigma_p)$ plane.
(Middle panels) the favored regions in the $(m_{\chi},\xi)$ plane.
The best fit points are also shown as asterisks.
(Bottom panels) the favored regions in the $(m_{\chi},m_{\phi})$ plane.
}
\end{figure}

Finally, we perform a combined $\chi^{2}$ analysis on the  results of
the three experiments CDMS-II-Si, SuperCDMS and LUX,
and
allow all the four parameters
$m_{\chi}$, $m_{\phi}$, $\bar\sigma_{p}$ and $\xi$ to vary.
The results of the allowed regions in the parameter space are shown
in \fig{fig:global}.
It can be seen that
after including the results of SuperCDMS and LUX,
in general
a light mediator with $m_{\phi} \lesssim 20$ MeV is favored by the data.
The allowed mass of DM particle is slightly reduced,
but still much larger than the case with contact interaction.
The value of isospin-violation parameter $\xi$ is found to be
i.e., $\xi= -0.70\pm 0.02$, very close to xenonphobic,
which is driven by the LUX data.

For type-I operators,
at $68\%$ CL  the allowed range of $m_{\chi}$ is $\sim 8-23$ GeV.
At $90\%$ CL, the allowed $m_{\chi}$ can reach 80 GeV.
For type-II operators,
at $68\%$ CL, the allowed range of $m_{\chi}$ is $\sim 6-13$ GeV.
At $90\%$ CL, the allowed $m_{\chi}$ can reach $\sim 20$ GeV.
Note that there is a local minimum at $\xi = -0.78$ and $m_{\chi} \sim 6\GeV$,
which corresponds to the ``Ge-phobic" DM~\cite{Gelmini:2014psa}.
A ``Ge-phobic" DM can maximally relax the constraint from SuperCDMS.
In the case of type-III operators,
the result is similar to the case of type-I operators.

\begin{figure}
\begin{center}
\subfloat[]{\includegraphics[width=0.5\textwidth]{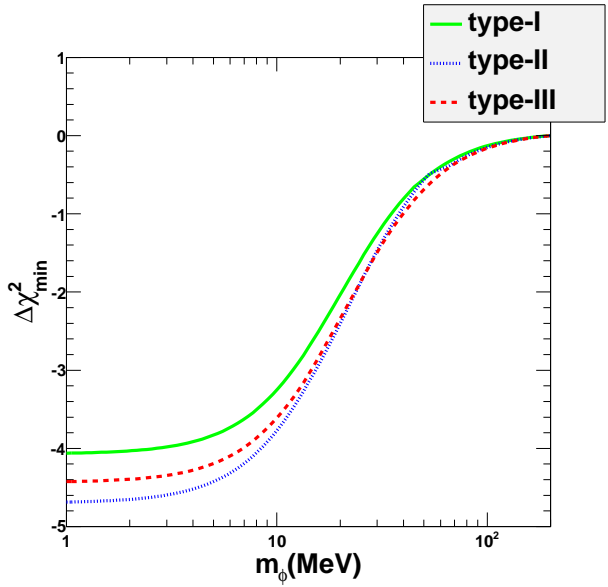}}

\end{center}
\caption{\label{fig:chi2_par}
The value of $\Delta\chi^2_{\text{min}}$ as a function of $m_\phi$, based on the global fits on the combination of CDMS-II-Si, LUX and SuperCDMS. The curves correspond to the operators of type-I (green solid), type-II (blue dotted) and type-III (red dashed).
}
\end{figure}

The presence of light mediator can help in
relaxing the tensions among the three experiments.
In \fig{fig:chi2_par},
we plot the quantity
$\Delta \chi_{\text{min}}^{2}=
\chi_{\text{min}}^{2}(m_{\phi})-\chi_{\text{min}}^{2}(m_{\phi}=200 \text{MeV})
$
which is the $\chi^{2}_{\text{min}}$ as a function of $m_{\phi}$ relative to
that at a large $m_{\phi}=200$ MeV.
The figure shows that for all the three type of operators,
the value of $\Delta \chi_{\text{min}}^{2}$ decreases for a decreasing $m_{\phi}$.
One sees that
the reduction of $\Delta \chi_{\text{min}}^{2}$ can be $\sim 4-5$.
%

\section{Conclusion\label{sec:Conclusion}}
In summary,
we have discussed the scenario where
the DM particle scatters off target nuclei through
the exchange of  a light mediator particle.
In this case,
the spectral feature of the recoil event rate depends on
the mass of the mediator  and type of interactions,
and can be  significantly different from
the case of contact-like interactions.
We have adopted an extended effective operator framework,
and discussed the momentum and velocity dependences of
the scattering cross sections for all the operators.
The interpretation of the current experimental data of
CDMS-II-Si, SuperCDMS and LUX, etc. has been discussed
with the focus on the determination of DM particle mass for
the relevant operators.
We have found  that in general
the allowed DM particle mass from the experimental data
increases significantly when the mediator particle becomes lighter.
For the data of CDMS-II Si,
the allowed DM particle mass  can reach
$\sim 50 (100)$ GeV at 68\% (90\%) CL 
depending on the type of operators,
and
the increasement saturates  when
the mediator mass is below $\mathcal{O}(10)$ MeV.
The upper limits from other experiments such as
SuperCDMS, CDMSlite, CDEX, XENON10/100, LUX, PandaX etc.
all become weaker at  high DM mass region.
In a combined analysis,
we have shown that the presence of a light mediator can
partially relax the tension in the current results of  CDMS-II-Si,
SuperCDMS and LUX.

\subsection*{Acknowledgments}
This work is supported in part by
the National Basic Research Program of China (973 Program) under Grants
No. 2010CB833000;
the National Nature Science Foundation of China (NSFC) under Grants
No. 10821504,
No. 11335012 and
No. 11475237.

\section* {Appendix}
\appendix
\section{Subdominate spin-independent operators}
In this section, we list the $G(q^{2},v)$ factors for the subdominant operators
$\mathcal{O}_{7}-\mathcal{O}_{9}$. 
They can be catalogued into the following types.

\noindent
{\bf Type-IV operator} $\mathcal{O}_{9}$ contributes to  the matrix elements of the form
\begin{align}
|M_{\chi N}(q^{2},v)|^{2}
&\propto
\frac{q^{4}}{(q^{2}+m_{\phi}^{2})^{2}},
\end{align}
The $G(q^{2},v)$ factor for this type of matrix element  is
\begin{align}
G_{4}(q^{2})=\frac{q^{4}/m_{\phi}^{4}}{I_{4}\left(
q^{2}_{\text{min}}/m^{2}_{\phi}, q_{\text{ref}}^{2}/m^{2}_{\phi}\right)
\left(1+q^{2}/m_{\phi}^{2}\right)^{2}
} ,
\end{align}
where
\begin{align}
I_{4}(a,b)
\equiv
\frac{1}{b-a}\int^{b}_{a} dt \frac{t^2}{(1+t)^{2}}
=
1- I_1(a,b)-2 I_2(a,b),
\end{align}
In the limit  of \eq{eq:heavylimit},
the function $I_{4}$ has an asymptotic form
$I_{4}\approx q_{\text{ref}}^{2}/(3 m_{\phi}^{2})$.

\noindent
{\bf Type-V operators}
$\mathcal{O}_{8}$ and  $\mathcal{O}_{9}$ also contribute to
the matrix elements of the form
\begin{align}
|M_{\chi N}(q^{2},v)|^{2}
&\propto
\frac{v_{\bot}^{2} q^{2}}{(q^{2}+m_{\phi}^{2})^{2}},
\end{align}
with
\begin{align}
G_{5}(q^{2},v)=\frac{v_{\bot}^{2}q^{2}/( v^{2}_{\text{ref}} m_{\phi}^{2}) }{I_{5}\left(
q^{2}_{\text{min}}/m^{2}_{\phi}, q_{\text{ref}}^{2}/m^{2}_{\phi}\right)
\left(1+q^{2}/m_{\phi}^{2}\right)^{2}
} ,
\end{align}
where $I_{5}(a,b) = I_{2}(a,b)-I_{4}(a,b)/b$.
In the limit  of \eq{eq:heavylimit},
the function $I_{5}$ has an asymptotic form
$I_{5}\approx q_{\text{ref}}^{2}/( 6 m_{\phi}^{2})$.

\noindent
{\bf Type-VI operators}
$\mathcal{O}_{7}$  contribute to
the matrix elements of the form
\begin{align}
|M_{\chi N}(q^{2},v)|^{2}
&\propto
\frac{v_{\bot}^{2} q^{4}}{(q^{2}+m_{\phi}^{2})^{2}},
\end{align}
For type-VI operators,
\begin{align}
G_{6}(q^{2},v)=\frac{v_{\bot}^{2}q^{4}/( v^{2}_{\text{ref}} m_{\phi}^{4}) }
{I_{6}\left(q^{2}_{\text{min}}/m^{2}_{\phi}, q_{\text{ref}}^{2}/m^{2}_{\phi}\right)
\left(1+q^{2}/m_{\phi}^{2}\right)^{2}
},
\end{align}
where
\begin{align}
I_{6}(a,b)
&\equiv
I_4(a,b)-\frac{1}{b(b-a)}\int^{b}_{a} dt \frac{t^3}{(1+t)^{2}} \nonumber \\
&=
\frac{ab(a+b)+(b-a)^2-3(a+b)-6}{2b(1+a)(1+b)}+\frac{3}{b(b-a)}\ln{\left(\frac{1+b}{1+a}\right)}.
\end{align}

\bibliographystyle{JHEP}
\bibliography{paper,misc}

\end{document}